\documentclass[journal,draftclsnofoot,onecolumn,12pt,twoside]{IEEEtranTCOM}

\normalsize

\ifCLASSINFOpdf

\else

\fi

\usepackage{amsmath}
\usepackage{wasysym}

\usepackage[ruled,vlined]{algorithm2e}
\usepackage{amssymb}
\usepackage{mathtools}
\usepackage{tikz}
\usepackage{pgf}
\usepackage{pgfplots}
\usepackage{cite}

\usetikzlibrary{spy}

\usetikzlibrary{external}

\definecolor{myblue}{rgb}{0.00000,0.44700,0.74100}
\definecolor{myred}{rgb}{0.85000,0.32500,0.09800}
\definecolor{mycolor3}{rgb}{0.92900,0.69400,0.12500}
\definecolor{mypurple}{rgb}{0.49400,0.18400,0.55600}
\definecolor{mygreen}{rgb}{0.46600,0.67400,0.18800}

\pgfplotsset
{colormap={InueJet}{rgb255(0cm)=(255,255,255) rgb255(0.25cm)=(218,236,255) rgb255(0.5cm)=(182,255,255) rgb255(0.75cm)=(200,255,200) rgb255(1.25cm)=(255,255,109) rgb255(1.75cm)=(255,163,72) rgb255(2.5cm)=(255,36,36) rgb255(3.5cm)=(128,0,0)}
}

\begin{document}

\title{Efficient Precoding Scheme for Dual-Polarization Multi-Soliton Spectral Amplitude Modulation}

\author{Alexander Span, Vahid Aref, Henning B\"ulow, and Stephan ten Brink
\thanks{A. Span and S. ten Brink are with the Institute of Telecommunications, University of Stuttgart, Stuttgart, Germany}
\thanks{V. Aref and H. B\"ulow are with Nokia Bell Labs, Stuttgart, Germany.}
}

\markboth{}%
{}

\maketitle

\vspace{-2cm}
\begin{abstract}
Spectral amplitude modulation for dual-polarization multi-soliton transmission is considered. We show, that spectral amplitudes become highly correlated during propagation along a noisy fiber link. Thus, joint equalization is generally needed at the receiver.

We propose a simple precoding scheme to almost remove these correlations. It significantly improves the detection performance even without any complex equalization. The spectral amplitudes are transformed into pairs of common and differential information part. We show, that the differential part is almost preserved along the link, even in the presence of noise. Thus, it can be detected directly from the received spectral amplitudes with high reliability. Using the precoding, disjoint detection of the common and the differential part causes only a small performance loss.

We analyze our precoding and verify its performance gain in Split-Step-Fourier simulations by comparing it to the conventional independent modulation of spectral amplitudes for first and second order solitons. 
\end{abstract}


\IEEEpeerreviewmaketitle

\section{Introduction}
\IEEEPARstart{A}{s} conventional optical transmission schemes suffer from nonlinear distortion, the fiber nonlinearity is commonly recognized as the limiting factor for the achievable data rate of optical communication systems~\cite{essiambre2010}. So far, nonlinearity has been considered as an undesirable effect to be mitigated. Thus, different nonlinearity compensation techniques such as digital backpropagation~\cite{ip2008} or optical phase conjugation~\cite{Chen2012} have been applied. 

Another approach is based on the Nonlinear Fourier Transform (NFT)~\cite{SergeiK.Turitsyn2017}. The Kerr-nonlinearity is treated as a fundamental property of the fiber channel and is taken into account for the transmission system design. For an ideal nonlinear optical fiber modeled by the standard Nonlinear Schroedinger Equation (NLSE), i.e. in the absence of noise and loss, the NFT linearizes the channel. It maps an optical time domain signal to a so-called nonlinear spectrum where the nonlinear crosstalk is absent. Despite the nonlinear behavior of pulses in time and linear frequency domain, the evolution equations in the nonlinear spectrum are simple, linear and thus, easily invertible at the receiver. This concept is also known as the Inverse Scattering Transform (IST) and motivates to encode data in the nonlinear spectrum.

The NFT is also applicable for dual polarization signals where the propagation is described by two coupled NLSEs, known as the Manakov system \cite{Manakov1974}.
The nonlinear spectrum consists of a continuous and a discrete part. The continuous part is the analogue to the conventional Fourier transform and describes the radiative pulse component, while the discrete spectrum describes the solitonic part. A special class of pulses, known as multi-solitons of order $N$ are described by the discrete part only, consisting of $N$ eigenvalues and $N$ corresponding spectral amplitudes.

Different ways of modulating the nonlinear spectrum have been studied and demonstrated experimentally for single and dual polarization.
For single polarization, the application of various classical QAM modulation formats has been investigated in the continuous nonlinear spectrum \cite{le2014nonlinear,le2015nonlinear,Le2016,prilepsky2014nonlinear,Le2017b,Le2017a}.  While first investigations of eigenvalue on-off keying date back almost three decades \cite{mollenauer2006solitons,Hasegawa1993}, the recent coherent technology allowed higher order modulation formats for eigenvalue  \cite{dong2015nonlinear,aref2016onoff,hari2016multieigenvalue} and spectral amplitude modulation\cite{aref2015experimental,buelow2016experimental,Geisler2016,buelow20167eigenvalues}.

To increase the spectral efficiency, both orthogonal polarizations need to be exploited~\cite{SergeiK.Turitsyn2017}. Polarization-division multiplexing for NFT-based transmission is described in \cite{goossens2017}. Dual polarization continuous spectrum modulation was investigated in \cite{Civelli2018} and polarization-division multiplexed eigenvalue modulation was introduced in \cite{maruta2015}. QPSK phase modulation of two eigenvalues has been experimentally demonstrated in \cite{Gaiarin2017,Gaiarin2018} for an EDFA fiber link up to $747\mathrm{km}$.

Although the concept of IST is applicable in practice, NFT-based transmission systems suffer from the non-ideality of the channel. When propagating along a noisy fiber link, the noise interacts nonlinearly with the signal. The eigenvalues as well as the spectral amplitudes of a soliton pulse get perturbed and especially become correlated. The evolution equations in the nonlinear spectrum are, however, only exact in the absence of noise and attenuation. Thus, the IST becomes less precise, effectively decreasing the detection performance and reducing the spectral efficiency.

In single polarization \cite{Gui2017,aref2016spectral} and in dual polarization \cite{Gaiarin2018a}, the detection performance can be improved by using minimum mean square error (MMSE) estimators based on the correlations between spectral amplitudes and eigenvalues. These schemes are, however, computationally complex with limited gain.

In this paper, we suggest an alternative simple solution that improves the detection performance without any complex equalization. Instead of exploiting the correlations for equalization, it applies a precoding such that the correlations are almost removed. 

We show that, when propagating along a noisy link, the perturbation on the spectral amplitudes is mainly caused and dominated by the eigenvalue fluctuations. Each pair of spectral amplitudes per eigenvalue shares a common, eigenvalue-dependent transformation term along the link. Thus, they undergo a similar transformation during the transmission, even in the presence of noise. Consequently, these spectral amplitudes become highly correlated and their differential information is almost preserved along the link.

We exploited this observation in \cite{Span2018} and introduced a precoding scheme for differential modulation. Here, we give a more detailed explanation and generalize this scheme to all available degrees of freedom offered by the spectral amplitudes. The spectral amplitudes are transformed into another set of quantities, denoted as differential and common parts. The differential part can be reliably detected directly from the received nonlinear spectrum without equalization, while the common part still needs to be equalized. However, the disjoint detection of the common and the differential part causes a smaller performance loss as compared to disjoint detection without the precoding.

The differential precoding is motivated by modeling the spectral amplitude perturbation along the link by a first order Markov model. The performance gain of the precoding scheme is shown in Split Step Fourier Method (SSFM) simulations and compared to the classical independent spectral amplitude modulation. We carry out simulations for first and second order solitons for transmission along an EDFA fiber link of $2988\mathrm{km}$.

The paper is outlined as follows: In Sec.~\ref{sec:prelim}, we briefly review the basics of the NFT and respective numerical algorithms. The perturbation of spectral amplitudes and their relation to the eigenvalue fluctuations is investigated in Sec.~\ref{sec:motivation}. Our precoding scheme for the differential modulation is presented in Sec.~\ref{sec:precoding}. Its detection performance gain is verified numerically and compared to conventional modulation schemes by simulation in Sec.\ref{sec:simulation}. The paper is concluded in Sec.~\ref{sec:conc}.

\section{Basics of Nonlinear Fourier Transform}\label{sec:prelim}

The normalized pulse propagation in two polarizations along an ideal, noiseless optical fiber is characterized by the following two coupled nonlinear Schroedinger Equations, known as the Manakov system. The signal vector ${\mathbf{q}(t,z)=\left(q_1(t,z),q_2(t,z)\right)^T}$ represents the pulse's components in the two orthogonal polarizations.

\begin{align}\label{eq:Manakov_system}
\frac{\partial \mathbf{q}(t,z)}{\partial z}+j\frac{\partial^2 \mathbf{q}(t,z)}{\partial t^2}+2j\left\|\mathbf{q}(t,z)\right\|^2 \mathbf{q}(t,z)=0	
\end{align}

The above equation is without physical units. The pulse evolution along the fiber is obtained by the following scaling:
\begin{align}\label{eq:Q_physical}
& Q_k\left(\tau,\ell\right)=\sqrt{P_0}\,\,\, q_k \left(\frac{\tau}{T_0},\frac{\ell}{L_0} \right)
\\
& \text{ with } P_0\cdot T_0^2=\frac{\left|\beta_2\right|}{\frac{8}{9}\gamma}, \quad L_0=\frac{2T_0^2}{\left|\beta_2\right|}, \quad k=1,2 \nonumber
\end{align}
where $P_0$ determines the physical power, $\beta_2<0$ and $\gamma$ are the chromatic dispersion and Kerr nonlinearity factor, respectively. $T_0$ determines the pulse duration and the propagation distance is scaled by $L_0$.

Similar to the single polarization case, the closed form solution of~\eqref{eq:Manakov_system} can be represented in the nonlinear spectrum defined by the extended Zakharov-Shabat system,
\begin{equation}\label{eq:ZS_Manakov}
\frac{\partial}{\partial t}\boldsymbol{\vartheta}
= \Lambda
\boldsymbol{\vartheta}
\end{equation}
\begin{equation*}
\boldsymbol{\vartheta} =\left( \begin{matrix} \vartheta_1(\lambda;t,z) \\ \vartheta_2(\lambda;t,z) \\ \vartheta_3(\lambda;t,z) \end{matrix} \right), \quad \Lambda=\left( \begin{matrix} -j\lambda & q_1(t,z) & q_2(t,z) \\ -q_1^*(t,z) & j\lambda & 0 \\ -q_2^*(t,z) & 0 & j\lambda \end{matrix} \right)	\nonumber
\end{equation*}
with the boundary condition
\begin{equation}
\boldsymbol{\vartheta}
\to
\left( \begin{matrix} 1 \\ 0 \\ 0 \end{matrix} \right) \exp(-j\lambda t)
\qquad \mathrm{for} \quad t\to -\infty.
\end{equation}
The nonlinear (Jost) coefficients are defined as.
\begin{align}\label{eq:spec_coeffs}
a(\lambda;z)&=\lim_{t\to \infty} \vartheta_1(\lambda;t,z) \exp(j\lambda t)
\\
b_1(\lambda;z)&=\lim_{t\to \infty} \vartheta_2(\lambda;t,z) \exp(-j\lambda t)
\\
b_2(\lambda;z)&=\lim_{t\to \infty} \vartheta_3(\lambda;t,z) \exp(-j\lambda t)
\end{align}

An important property of these nonlinear coefficients is the simple description of the pulse propagation. When optical pulses evolve along the \emph{ideal} fiber in $z$, the nonlinear coefficients are transformed as
\begin{align}\label{eq:b_evol}
a(\lambda;z_0+z) & = a(\lambda;z_0) = a(\lambda)	\nonumber
\\
b_{i}(\lambda;z_0+z) & =b_{i}(\lambda;z_0) \exp(-4j\lambda^2 z)
\end{align}

For an ideal transmission, $a(\lambda)$ is thus unchanged and $b(\lambda;z)$ is transformed linearly. At every position $z$ along the link, a time domain pulse is equivalently described in the nonlinear spectrum by:
\begin{itemize}

\item Discrete spectrum: $\{\lambda_k,b_1(\lambda_k;z),b_2(\lambda_k;z)\}$ where the eigenvalues $\lambda_k=\omega_k+j\sigma_k \in \mathbb{C}^+$ ($\omega_k,\sigma_k \in \mathbb{R}$) are the roots of $a(\lambda)$.

\item Continuous spectrum: $\{b_1(\lambda;z),b_2(\lambda;z)\}$, for $\lambda \in \mathbb{R}$.

\end{itemize}

Note that, as $a(\lambda)$ does not change during ideal propagation in $z$, the eigenvalues are also preserved, $\lambda_k(z)=\lambda_k$.

The single polarization case is obtained by setting $q_2(t)$ to zero. Thus, in the nonlinear spectrum, $b_2(\lambda;z)$ is the additional degrees of freedom offered by the second polarization. 

There are various ways of computing the nonlinear spectrum by numerically solving~\eqref{eq:ZS_Manakov}. We use the Forward-Backward method~\cite{aref2016control} and extend it to two polarizations (as in \cite{Gaiarin2018}). This method achieves the highest accuracy and numerical stability in the computation of the spectral coefficients.

The Inverse Nonlinear Fourier Transform (INFT) is the operation to generate the time domain signal $\left(q_1(t), q_2(t)\right)^T$ given a nonlinear spectrum. The special class of multi-soliton pulses is determined by the discrete part of the spectrum only and the continuous spectrum is equal to zero, $b_1(\lambda;z)=b_2(\lambda;z)=0$ for $\lambda \in \mathbb{R}$. In that case, the Darboux transformation can be applied to generate multi-soliton pulses from the given eigenvalues and discrete nonlinear spectral coefficients. We use Alg.~\ref{alg:DT2} to perform the Darboux transform~\cite{Wright2003}.

\begin{algorithm}

\SetKwInOut{Input}{Input}

\SetKwInOut{Output}{Output}

\Input{Discrete Spectrum $\{\lambda_k, \, b_1(\lambda_k), \, b_2(\lambda_k)\}$; $k=1,\dots,N$}

\Output{Dual polarization soliton waveform $\left( q_1(t), q_2(t) \right) ^T$}

\BlankLine

\Begin{

\For{$k\leftarrow 1$ \KwTo $N$}{
\begin{flalign*}
& \boldsymbol{\vartheta}_k^{(0)}=\left( \begin{matrix} \vartheta_{1,k}^{(0)} \\ \vartheta_{2,k}^{(0)} \\ \vartheta_{3,k}^{(0)} \end{matrix} \right)
\leftarrow
\left( \begin{matrix} \exp(-j\lambda_k t) \\ b_1(\lambda_k) \exp(j\lambda_k t) \\ b_2(\lambda_k) \exp(j\lambda_k t) \end{matrix} \right) &
\end{flalign*}
}

$q_{1/2}^{(0)}(t) \longleftarrow 0$\;

\For{$k\leftarrow 1$ \KwTo $N$}{

\vspace*{-.5cm}
\begin{flalign}\label{eq:sig_update}
& q_{1/2}^{(k)}(t) \leftarrow q_{1/2}^{(k-1)}(t) +  \frac{2j\left(\lambda_k^* - \lambda_k\right) \left(\frac{\vartheta_{2/3,k}^{(k-1)}}{\vartheta_{1,k}^{(k-1)}}\right)^*}{1+\left|\frac{\vartheta_{2,k}^{(k-1)}}{\vartheta_{1,k}^{(k-1)}}\right|^2+\left|\frac{\vartheta_{3,k}^{(k-1)}}{\vartheta_{1,k}^{(k-1)}}\right|^2} &
\end{flalign}

\For{$m\leftarrow k+1$ \KwTo $N$}{

\begin{flalign*}
& \boldsymbol{H} \leftarrow \left(\begin{matrix} \vartheta_{1,k}^{(k-1)} & \vartheta_{2,k}^{*(k-1)} & \vartheta_{3,k}^{*(k-1)} \\ \vartheta_{2,k}^{(k-1)} & -\vartheta_{1,k}^{*,(k-1)} & 0 \\ \vartheta_{3,k}^{(k-1)} & 0 & -\vartheta_{1,k}^{*,(k-1)} \end{matrix}\right) &
\end{flalign*}

\begin{flalign*}
& \boldsymbol{M} \leftarrow \boldsymbol{H}  \left( \begin{matrix} \lambda_k & 0 & 0 \\ 0 & \lambda_k^* & 0 \\ 0 & 0 & \lambda_k^* \end{matrix} \right)   \boldsymbol{H}^{-1} &
\end{flalign*}
\begin{flalign*}
& \boldsymbol{\vartheta}_m^{(k)} \leftarrow \left( \lambda_m \boldsymbol{I} - \boldsymbol{M} \right)  \boldsymbol{\vartheta}_m^{(k-1)} &
\end{flalign*}

}

}

}

\caption{INFT via Darboux transform algorithm\label{alg:DT2}}

\end{algorithm}

\section{Motivation - Transformation of $b$}\label{sec:motivation}

The simple transformation \eqref{eq:b_evol} motivates to encode the data in the nonlinear spectrum.  After propagation, the information can be recovered at the receiver by simply inverting this transformation. However, \eqref{eq:b_evol} is only exact for pulse propagation along an \emph{ideal} fiber. In the presence of noise, the eigenvalues $\lambda_k$ and the spectral coefficients $b_1(\lambda_k;z)$, $b_2(\lambda_k;z)$ are perturbed. The transformation \eqref{eq:b_evol} is no longer valid and the precise noise model for the nonlinear spectrum is not yet available. In~\cite{zhang2015spectral}, a Markov model was introduced to approximate the transformation of the nonlinear spectrum along a noisy link. Based on this model, estimation methods for the nonlinear spectrum in single polarization were studied in \cite{Gui2017,aref2016spectral}.

\begin{figure}
\centering
\includegraphics[scale=0.8]{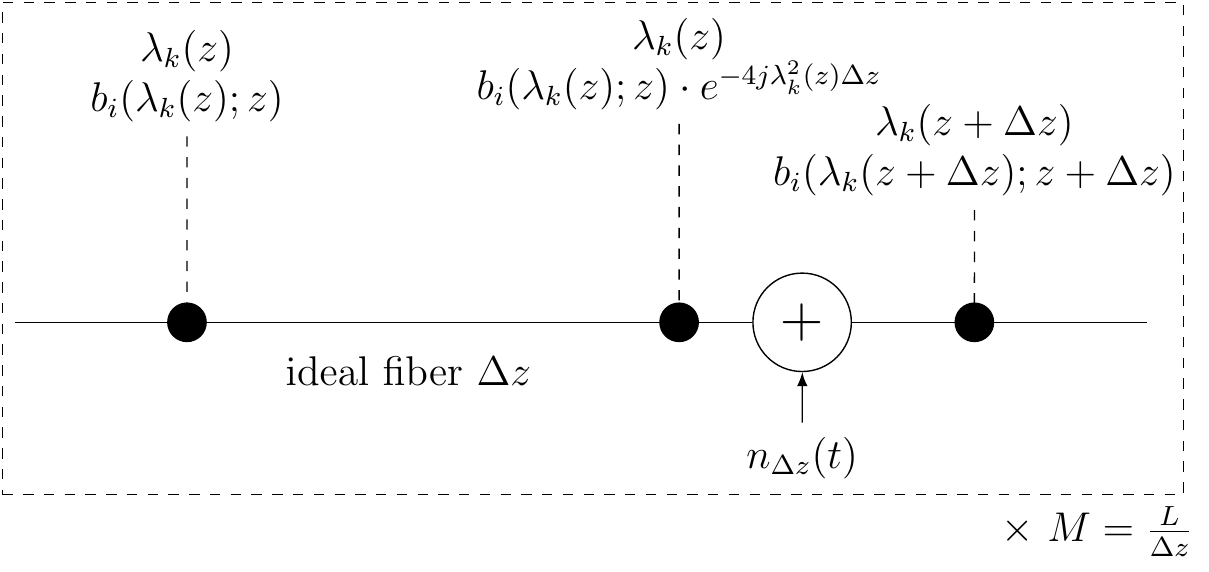}
\caption{Model for Raman amplified fiber link: Concatenation of ideal fiber link of length $\Delta z \to 0$ followed by additive noise}
\label{fig:distRaman_model}
\end{figure}

In the following, we apply the Markov model from~\cite{zhang2015spectral} to the dual polarization case. For that, let us first consider (ideal) distributed Raman amplification, for which the Manakov system is given as
\begin{align}\label{eq:Manakov_system_noise}
\frac{\partial \mathbf{q}(t,z)}{\partial z}+j\frac{\partial^2 \mathbf{q}(t,z)}{\partial t^2}+2j\left\|\mathbf{q}(t,z)\right\|^2 \mathbf{q}(t,z)=n(t,z)
\end{align}
The noise term $n(t,z)$ is AWGN with noise power spectral density (PSD) $N_0=n_\mathrm{sp} \alpha h \nu$ (attenuation coefficient $\alpha$, Planck constant $h$, carrier frequency $\nu$). The noise is distributed along the fiber and attenuation is perfectly compensated. Such a link of length $L$ can be modeled by the concatenation of $M=L/\Delta z$ short ideal fiber segments of length $\Delta z \to 0$, each followed by an additive noise injection $n_{\Delta z}(t)$ with noise PSD $N_0 \Delta z$ as shown in Fig.~\ref{fig:distRaman_model}. We assume that the noise PSD is so small that it does not generate new eigenvalues and the existing eigenvalues are only slightly perturbed.

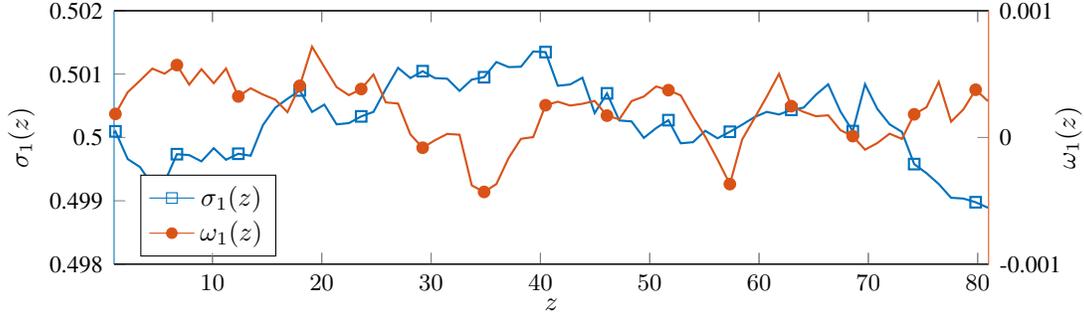
\begin{figure}
\center
\begin{tikzpicture}[baseline=(current axis.south)]
\begin{axis}[axis y line*=left,ticklabel style={font=\footnotesize},width=0.8\columnwidth,height=0.3\columnwidth,xmin=1,xmax=81,ymin=0.498,ymax=0.502,label style={font=\small},ylabel={$\sigma_1(z)$},y label style={yshift=0em},xlabel={$z$},x label style={yshift=0.5em},ytick={0.498,0.499,0.5,0.501,0.502},yticklabels={0.498,0.499,0.5,0.501,0.502},legend style={font=\small},title style={font=\small},y axis line style={myblue},legend pos=south west]

\addplot[forget plot,thick,myblue,mark=square,mark repeat=5,width=\linewidth]table {1Sol_ldvImz-1.dat};

\addlegendimage{myblue,mark=square}
\addlegendentry{$\sigma_1(z)$}    
\addlegendimage{myred,mark=*}
\addlegendentry{$\omega_1(z)$}
   
\end{axis}
\begin{axis}[axis y line*=right,ticklabel style={font=\footnotesize},width=0.8\columnwidth,height=0.3\columnwidth,xmin=1,xmax=81,ymin=-0.001,ymax=0.001, label style={font=\small},y label style={yshift=-33em},ylabel={$\omega_1(z)$}, ytick={-0.001,0,0.001},yticklabels={-0.001,0,0.001},yticklabel pos=right,legend style={font=\small},xtick={},xticklabels={}title style={font=\small},y axis line style={myred},scaled y ticks={real:10},scaled ticks=false]  
  
\addplot[forget plot,thick,myred,mark=*,mark repeat=5,width=\linewidth]table {1Sol_ldvRez-1.dat};

\end{axis}
\end{tikzpicture}

\caption{Eigenvalue trajectory $\lambda_1(z)=\omega_1(z)+j\sigma_1(z)$ of exemplary first order soliton with ${\lambda_1(0)=0.5j}$ when propagating along a noisy fiber link.}

\label{fig:oneSoliton_EV_evolution_ex}
\end{figure}

Considering the ideal evolution equation \eqref{eq:b_evol} in a noisy fiber link, we identify two perturbation effects. First, at each noise injection point, eigenvalues are shifted randomly in the complex plane, leading to a random walk trajectory $\lambda_k(z)$ along the link. Then, the spectral coefficient transformation along the small ideal fiber segment of length $\Delta z$ in our model is $b_i(\lambda_k(z);z)\exp(-4j\lambda_k^2(z)\Delta z)$. This transformation follows \eqref{eq:b_evol} piecewise, however $\lambda_k$ is replaced with the perturbed eigenvalues $\lambda_k(z)$ at the respective position. Fig.~\ref{fig:oneSoliton_EV_evolution_ex} shows an example of an eigenvalue trajectory for a first order soliton with initial $\lambda_1(0)=0.5j$.

The second perturbation effect is that the additive noise does also instantaneously perturb the spectral coefficient. As a result, the transformation of the spectral coefficients at position $z$ after propagating a small distance $\Delta z \to 0$ as in Fig.~\ref{fig:distRaman_model} becomes

\begin{equation}\label{eq:b_evol_dz}
b_i\left(\lambda_k(z+\Delta z);z+\Delta z\right) 
 = b_i(\lambda_k(z);z) \exp\left(-4j\lambda_k^2(z) \Delta z \right) \underbrace{(1+\varepsilon_i(z+\Delta z))}_{\tilde{\varepsilon}_{k,i}(z+\Delta z)} 
\end{equation}
where $\varepsilon_i(z)$ corresponds to the error term perturbing the spectral coefficient. As a result, the transformation of the spectral coefficients $b_i(\lambda_k(0);0)$ from the transmitter at $z=0$ to some position along the link $z=L$ follows by concatenation of ${M=L/\Delta z}$ small steps with $\Delta z \to 0$: 
\begin{align}\label{eq:b_evol_MM}
b_{i}(\lambda_k(L);L)	\nonumber
& \approx  b_{i}(\lambda_k(0);0) \prod_{m=0}^{M-1}  \exp\left(-4j \lambda_k^2(m \, \Delta z) \Delta z\right) \tilde{\varepsilon}_{k,i}((m+1)\cdot \Delta z)	\nonumber
\\
& =  b_{i}(\lambda_k(0);0) \exp\left(-4j \Delta z \sum_{m=0}^{M-1}\lambda_k^2(m\cdot \Delta z) \right) \prod_{m=1}^{M} \tilde{\varepsilon}_{k,i}(m \, \Delta z)	\nonumber
\\
& = b_{i}(\lambda_k(0);0) \exp\left(-4j \int_0^L \lambda_k^2(z) \partial z \right) \prod_{m=1}^{M} \tilde{\varepsilon}_{k,i}(m \, \Delta z)
\end{align}
Note that in a noiseless scenario, eigenvalues do not change and ${\tilde{\varepsilon}_{k,i}=1}$. Then \eqref{eq:b_evol_MM} becomes identical to \eqref{eq:b_evol}.

When detecting the received $b_i(\lambda_k(L);L)$ at position $z=L$, one needs to invert the transformation \eqref{eq:b_evol_MM} in order to recover the transmitted $b_i(\lambda_k(0);0)$. Let us now show that the error in \eqref{eq:b_evol_MM} due to the eigenvalue fluctuations $\lambda_k(z)$ dominates compared to the error from $\tilde{\varepsilon}_{k,i}(z)$. To do this, we consider two types of equalization. First, a genie-aided equalizer (GAE) which knows the eigenvalue trajectory $\lambda_k(z)$ along the link. The genie-aided equalizer is
\begin{align}
\hat{b}_{i,\mathrm{GA}}(\lambda_k) & = b_{i}(\lambda_k(L);L) \exp\left(4j \int_0^L \lambda_k^2(z) \partial z\right) \label{eq:ideal_inv}
\\
& = b_{i}(\lambda_k(0);0) \prod_{m=1}^{M} \tilde{\varepsilon}_{k,i}(m\cdot \Delta z) \label{eq:ideal_inv_error}
\end{align}
where \eqref{eq:ideal_inv_error} follows from \eqref{eq:b_evol_MM}. It totally removes the error arising from the eigenvalue fluctuations and only the perturbations directly acting on the spectral coefficients $\tilde{\varepsilon}_{k,i}(z)$ remain. This is of course impractical as the eigenvalues are not known during propagation except at both ends of the fiber at the transmitter $\lambda_k(0)$ and at the receiver $\lambda_k(L)$.

Then we consider another heuristic equalizer, the mean backrotation (MBR). This equalizer has a performance close to the MMSE equalization for the first order soliton~\cite{aref2016spectral}. The MBR \eqref{eq:practical_inv} estimates ${\int_0^L \lambda_k^2(z) \partial z \approx 0.5 \left(\lambda_k^2(0)+\lambda_k^2(L)\right)L}$. Inserting the transformation \eqref{eq:b_evol_MM}, one can observe the estimate being a perturbed version of the transmitted $b_i(\lambda_k(0);0)$ as in \eqref{eq:practical_inv_error}.
\begin{align}
\hat{b}_{i,\mathrm{MBR}}(\lambda_k)& =b_{i}(\lambda_k(L);L) \exp\left(4j \frac{\lambda_k^2(0)+\lambda_k^2(L)}{2} L \right)	\label{eq:practical_inv}
\\
&=b_i(\lambda_k(0);0) \frac{\exp\left(-4j\int_0^L \lambda_k^2(z)\partial z\right)}{\exp\left(-4j \frac{\lambda_k^2(0)+\lambda_k^2(L)}{2}L\right)}\prod_{m=1}^{M} \tilde{\varepsilon}_{k,i}(m\cdot \Delta z)	\label{eq:practical_inv_error}
\end{align}

We exemplarily consider the transmission of a single first order soliton with initial $\lambda_1(0)=0.5j$ and ${b_{1/2}(\lambda_1(0),0)=1/\sqrt{2}}$ along a noisy, ideally amplified fiber link. We assume $n_\mathrm{sp}=1.1$ and carrier frequency $\nu=193.55\, \mathrm{THz}$. The time scaling factor is chosen as $T_0=20\mathrm{ps}$ such that the first order soliton pulse duration (containing $99.99\%$ of the total soliton energy) becomes $T\approx 0.2\mathrm{ns}$. Except applying ideal distributed Raman amplification, the remaining fiber parameters and simulation setup are as described later in Sec.\ref{sec:simulation}. 

\begin{figure}
\center
\begin{tikzpicture}[baseline=(current axis.south)]
\begin{axis}[ticklabel style={font=\normalsize},width=0.8\columnwidth,height=0.25\columnwidth,xmin=0,xmax=81,
ymin=-0.06,ymax=0.28,y label style={yshift=0.25em},label style={font=\normalsize},ylabel={$\arg \, \frac{\hat{b}_1(\lambda_1)}{b_1(\lambda_1(0),0)}$},title style={font=\normalsize},legend pos=north west,view={145}{20},legend columns=2]

    \addplot[myblue,forget plot,mark=square,mark repeat=10,mark phase=5,thick]table  {1Sol_b1z_phase_naiveIST-2.dat};      
    \addplot[mygreen,forget plot,thick,mark=square*,mark repeat=10]table  {1Sol_b1z_phase_idealIST-2.dat};

\addlegendimage{myblue,mark=square}
\addlegendentry{MBR}    
\addlegendimage{mygreen,mark=square*}
\addlegendentry{GAE}

\end{axis}

\end{tikzpicture}

\begin{tikzpicture}[baseline=(current axis.south)]
\begin{axis}[ticklabel style={font=\normalsize},width=0.8\columnwidth,height=0.25\columnwidth,xmin=0,xmax=81,
ymin=0.95,ymax=1.1,y label style={yshift=0.25em},label style={font=\normalsize},ylabel={$\frac{|\hat{b}_1(\lambda_1)|}{|b_1(\lambda_1(0),0)|}$},title style={font=\normalsize},legend pos=north west,view={145}{20}]

    \addplot[myblue,forget plot,thick,mark=square,mark repeat=10,mark phase=5]table {1Sol_b1z_abs_naiveIST-2.dat};      
    \addplot[mygreen,forget plot,thick,mark=square*,mark repeat=10]table  {1Sol_b1z_abs_idealIST-2.dat};

\end{axis}

\end{tikzpicture}

\begin{tikzpicture}[baseline=(current axis.south)]
\begin{axis}[ticklabel style={font=\normalsize},width=0.8\columnwidth,height=0.25\columnwidth,xmin=0,xmax=81,
ymin=-0.06,ymax=0.28,y label style={yshift=0.25em},label style={font=\normalsize},ylabel={$\arg \, \frac{\hat{b}_2(\lambda_1)}{b_2(\lambda_1(0),0)}$},title style={font=\normalsize},legend pos=north west,view={145}{20},legend columns=2]

    \addplot[myred,forget plot,thick,mark=o,mark repeat=10,mark phase=5]table {1Sol_b2z_phase_naiveIST-2.dat};      
    \addplot[mygreen,forget plot,thick,mark=*,mark repeat=10]table {1Sol_b2z_phase_idealIST-2.dat};

\addlegendimage{myred,mark=o}
\addlegendentry{MBR}    
\addlegendimage{mygreen,mark=*}
\addlegendentry{GAE}

\end{axis}

\end{tikzpicture}

\begin{tikzpicture}[baseline=(current axis.south)]
\begin{axis}[ticklabel style={font=\normalsize},width=0.8\columnwidth,height=0.25\columnwidth,xmin=0,xmax=81,
ymin=0.95,ymax=1.1,y label style={yshift=0.25em},label style={font=\normalsize},ylabel={$\frac{|\hat{b}_2(\lambda_1)|}{|b_2(\lambda_1(0),0)|}$},xlabel={$L$},title style={font=\normalsize},legend pos=north west,view={145}{20}]

    \addplot[myred,forget plot,thick,mark=o,mark repeat=10,mark phase=5]table  {1Sol_b2z_abs_naiveIST-2.dat};      
    \addplot[mygreen,forget plot,thick,mark=*,mark repeat=10]table  {1Sol_b2z_abs_idealIST-2.dat};

\end{axis}

\end{tikzpicture}
\caption{Estimation error of $\arg \, \hat{b}_i(\lambda_k)$ and $|\hat{b}_i(\lambda_k)|$ at position $L$ along the fiber for either GAE according to \eqref{eq:ideal_inv} or MBR according to \eqref{eq:practical_inv}. A first order soliton with $\lambda_1(0)=0.5j$ and $b_1(\lambda_1(0),0)=b_2(\lambda_1(0),0)=1/\sqrt{2}$ is transmitted. $L$ is in normalized units with the total propagation distance $L=80.9$ corresponding to $L\cdot L_0=2988\mathrm{km}$.}
\label{fig:1Sol_inversion}
\end{figure}
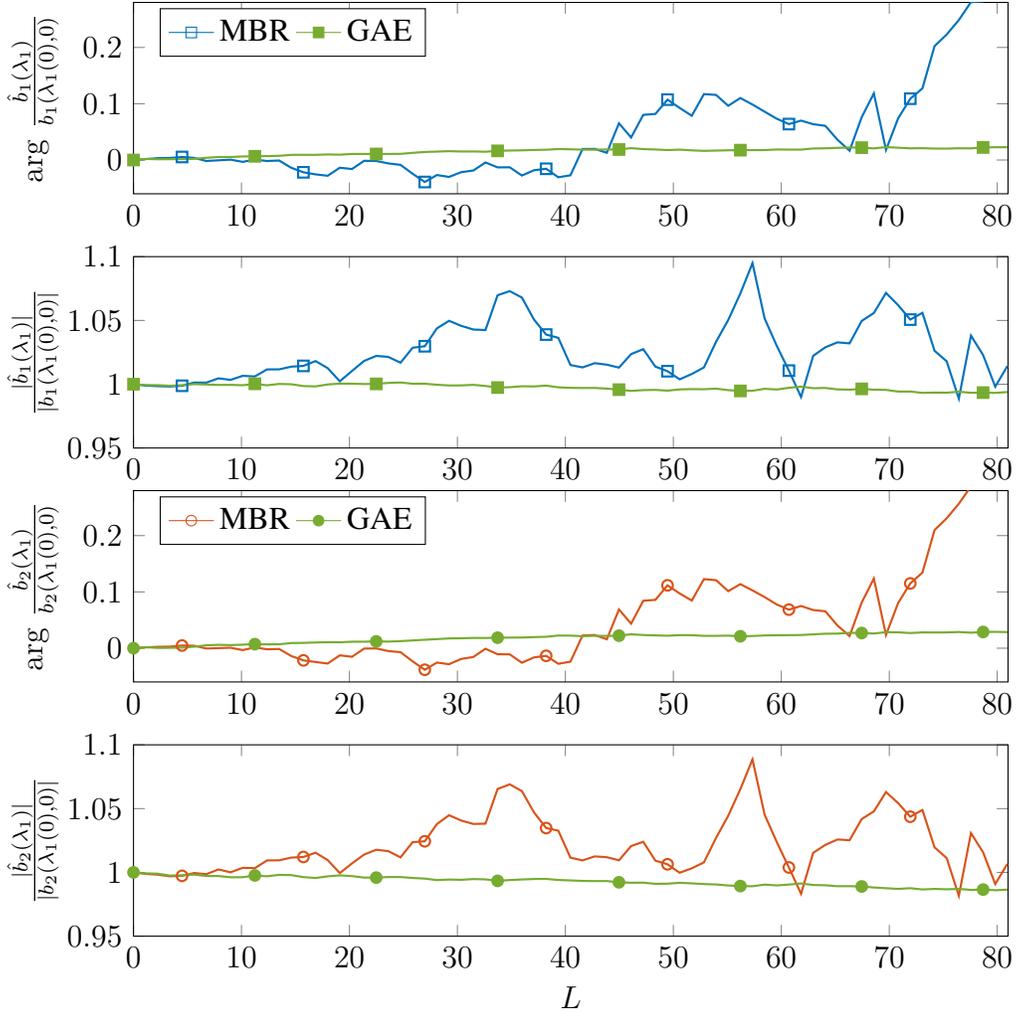

The estimation error $\hat{b}_i(\lambda_1)/b_i(\lambda_1(0),0)$ is shown in Fig.~\ref{fig:1Sol_inversion} when $\hat{b}_i(\lambda_1)$ is estimated either by the MBR \eqref{eq:practical_inv} or by the GAE \eqref{eq:ideal_inv} at different positions along the link $z=L$. The errors are separately shown for phase and magnitude of $\hat{b}_1(\lambda_1)$ and $\hat{b}_2(\lambda_1)$. It is visible in Fig.~\ref{fig:1Sol_inversion}, that the estimation error of $\hat{b}_{1,\mathrm{MBR}}(\lambda_k)$ and $\hat{b}_{2,\mathrm{MBR}}(\lambda_k)$, marked by $\square$/$\Circle$, are strongly correlated. This indicates that the errors are dominated by the eigenvalue fluctuations. Indeed, applying the GAE \eqref{eq:ideal_inv} removes this contribution and leads to the $\blacksquare$/$\CIRCLE$-curves. The remaining error is negligible and the transmitted $b_i(\lambda_k(0),0)$ can be recovered almost perfectly.

\begin{figure}
\center
\begin{tikzpicture}[baseline=(current axis.south)]
\begin{axis}[ticklabel style={font=\normalsize},width=0.8\columnwidth,height=0.3\columnwidth,xmin=0,xmax=81,
ymin=0.000008,ymax=0.03,y label style={yshift=0.35em},ymode=log,label style={font=\normalsize},xlabel={$L$},ylabel={$\mathrm{Var}\left(\arg\, \frac{\hat{b}_i(\lambda_1)}{b_i(\lambda_1(0),0)}\right)$},title style={font=\normalsize},legend pos=south east,view={145}{20},legend style={font=\small},legend columns=2]

    \addplot[myblue,forget plot,thick,mark=square,mark repeat=10]table [x expr=\thisrowno{0}*5.5225,y expr=\thisrowno{1}] {1Sol_b1PhaseEVM_naiveIST.dat};      
    \addplot[mygreen,forget plot,thick,mark=square*,mark repeat=10]table [x expr=\thisrowno{0}*5.5225,y expr=\thisrowno{1}] {1Sol_b1PhaseEVM_idealIST.dat};
    \addplot[myred,forget plot,thick,mark=o,mark repeat=10,mark phase=5]table [x expr=\thisrowno{0}*5.5225,y expr=\thisrowno{1}] {1Sol_b2PhaseEVM_naiveIST.dat};      
    \addplot[mygreen,forget plot,thick,mark=*,mark repeat=10,mark phase=5]table [x expr=\thisrowno{0}*5.5225,y expr=\thisrowno{1}] {1Sol_b2PhaseEVM_idealIST.dat};

\addlegendimage{myblue,mark=square}
\addlegendentry{$\hat{b}_{1,\mathrm{MBR}}$}    
\addlegendimage{myred,mark=o}
\addlegendentry{$\hat{b}_{2,\mathrm{MBR}}$}
\addlegendimage{mygreen,mark=square*}
\addlegendentry{$\hat{b}_{1,\mathrm{GAE}}$} 
\addlegendimage{mygreen,mark=*}
\addlegendentry{$\hat{b}_{2,\mathrm{GAE}}$}

\end{axis}
\begin{axis}[ticklabel style={font=\normalsize},width=0.8\columnwidth,height=0.3\columnwidth,xmin=0,xmax=81,
ymin=0.000008,ymax=0.03,ymode=log,label style={font=\normalsize},legend pos=north west,view={145}{20},xticklabels={,,},yticklabels={,,},legend style={font=\small},legend columns=1]

    \addplot[black,forget plot,thick,dashed]table [x expr=\thisrowno{0}*5.5225,y expr=\thisrowno{1}]{mean_squared_naive_minus_ideal_BR_error_1Sol.dat};

\addlegendimage{black,dashed}
\addlegendentry{$\overline{\Delta \varphi^2}$}

\end{axis}

\end{tikzpicture}
\caption{First order soliton QPSK modulation: Phase variance of estimated $\hat{b}_i(\lambda_k)$ along the link for either applying MBR \eqref{eq:practical_inv} or GAE \eqref{eq:ideal_inv}. Additionally, the mean squared phase difference between MBR and GAE \eqref{eq:channel_inv_error}. $L$ is in normalized units with the total propagation distance $L=80.9$ corresponding to $L\cdot L_0=2988\mathrm{km}$.}
\label{fig:phase_variance_naive_vs_ideal_inv}
\end{figure}
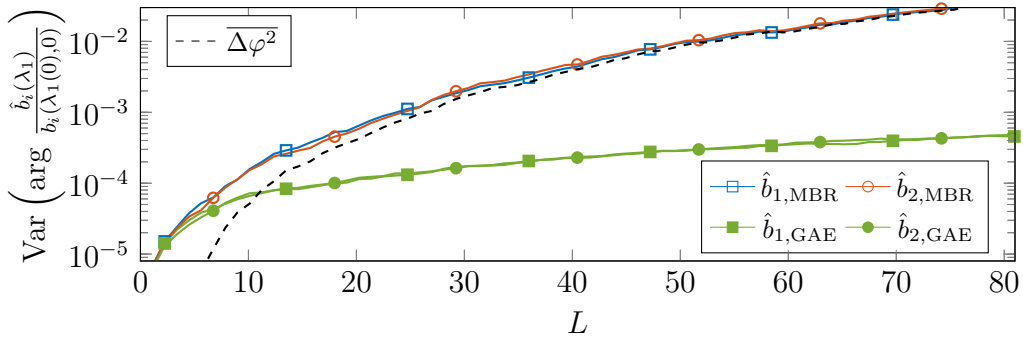

Fig.~\ref{fig:phase_variance_naive_vs_ideal_inv} illustrates the average estimation error for the MBR and the GAE when a train of first order solitons is transmitted along the link. Each $b_i(\lambda_1(0);0)$ is modulated independently using a QPSK format. The phase error variance is shown for both equalization methods. It can be observed that the GAE significantly reduces the phase variance. We also show the mean squared phase error arising from the unknown eigenvalue trajectory from \eqref{eq:practical_inv_error} as

\begin{equation}\label{eq:channel_inv_error}
\overline{\Delta \varphi^2}=\mathrm{E}\left[\arg\left\{ \frac{\exp\left(4j\int_0^L \lambda_k^2(z) \partial z \right)}
{\exp\left(4j \frac{\lambda_1(0)^2+\lambda_1^2(L)}{2} L\right)}\right\}^2\right].
\end{equation}

This phase error is shown by the dashed line in Fig.~\ref{fig:phase_variance_naive_vs_ideal_inv}. It can be observed, that the phase error arising from the eigenvalue fluctuation term \eqref{eq:channel_inv_error} eventually converges to the total phase variance of $\hat{b}_i(\lambda_1)/b_i(\lambda_1(0),0)$ when the transmission distance increases. It implies that the transformation perturbation on $b_i(\lambda_k(z),z)$ is dominated by the fluctuations of the eigenvalues compared to the small error terms $\prod_{m=1}^{M} \tilde{\varepsilon}_{k,i}(m\cdot \Delta z)$, if the link length exceeds a certain propagation distance. This is the reason for the high correlation between the errors in $\hat{b}_{1,\mathrm{MBR}}(\lambda_k)$ and $\hat{b}_{2,\mathrm{MBR}}(\lambda_k)$. The correlation can be seen in Fig.~\ref{fig:1Sol_inversion} where the error terms in phase and magnitude follow almost the same trajectory.

\section{Differential Precoding}\label{sec:precoding}

Let us consider the ratio of the spectral coefficients at position $L$: 
\begin{align}
& \frac{b_2(\lambda_k(L),L)}{b_1(\lambda_k(L),L)} 			  \label{eq:spec_amp_ratio}
\\
 = & \frac{b_{2}(\lambda_k(0);0) e^{-4j \int_0^L \lambda_k^2(z) \partial z } \prod_{m=1}^{M} \tilde{\varepsilon}_{k,2}(m \, \Delta z)}{b_{1}(\lambda_k(0);0) e^{-4j \int_0^L \lambda_k^2(z) \partial z } \prod_{m=1}^{M} \tilde{\varepsilon}_{k,1}(m \, \Delta z)}\label{eq:spec_amp_ratio_z}
\\
\approx & \frac{b_2(\lambda_k(0),0)}{b_1(\lambda_k(0),0)}	\label{eq:b1b2_relation}
\end{align}

where \eqref{eq:spec_amp_ratio_z} is obtained from \eqref{eq:b_evol_MM}. We have \eqref{eq:b1b2_relation} as $\prod_{m=1}^{M} \tilde{\varepsilon}_{k,i}(m\cdot \Delta z) \approx 1$.

\begin{figure}
\hspace{-5mm}
\includegraphics[scale=0.65]{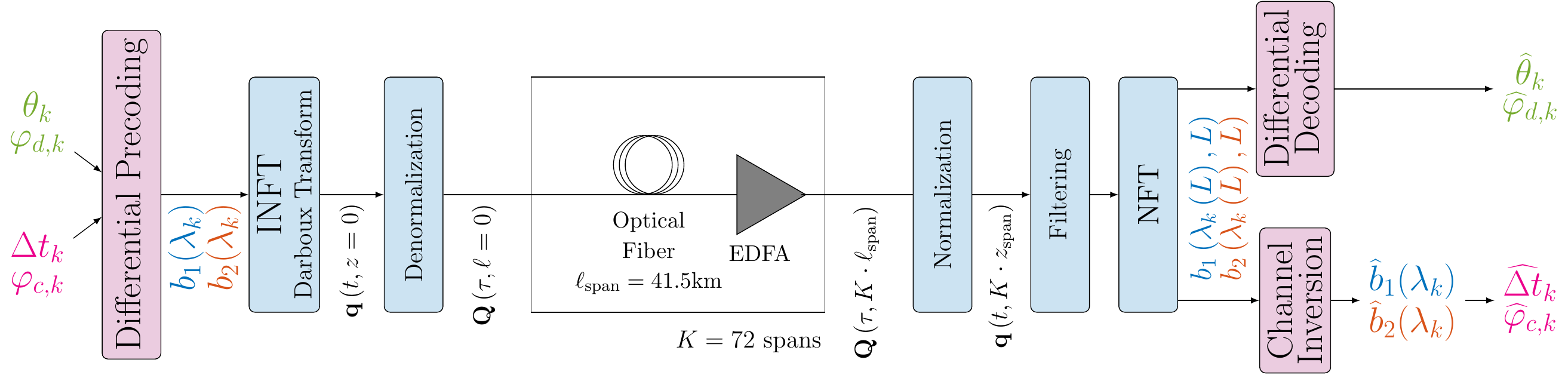}
\caption{Transmission simulation setup for differential spectral coefficient modulation of dual polarization multi-solitons}
\label{fig:trans_setup}
\end{figure}

As a result, the ratio \eqref{eq:b1b2_relation} remains approximately unchanged during the transmission and it is thus almost invariant to the eigenvalue fluctuations. This motivates to encode data differentially between the pairs of spectral coefficients. The transmission system with such an encoding and decoding scheme is shown in Fig.~\ref{fig:trans_setup} and explained in the following. For convenience, we denote the modulated spectral coefficients at the transmitter as $b_i(\lambda_k) \coloneqq b_i(\lambda_k(0),0)$.

\subsection{Encoding}\label{sec:Encoding}

Consider the eigenvalue $\lambda_k=j\sigma_k+\omega_k$. Instead of modulating directly $b_1(\lambda_k)$ and $b_2(\lambda_k)$, we can alternatively modulate the quantities $\Delta t_k$, $\theta_k$, $\varphi_{c,k}$ and $\varphi_{d,k}$ and map them to the spectral coefficients as
\begin{align}
b_1(\lambda_k) & =|b_1(\lambda_k)| \exp(j\varphi_{1,k})	\label{eq:b1_com_mapping}
\\
& =\exp(2\sigma_k \Delta t_k) |\cos(\theta_k)| e^{j\varphi_{c,k}}	\label{eq:b1_diff_mapping}
\\
b_2(\lambda_k) & =|b_2(\lambda_k)| \exp(j\varphi_{2,k})	\label{eq:b2_com_mapping}
\\
& =\exp(2\sigma_k \Delta t_k) |\sin(\theta_k)| e^{j\varphi_{c,k}+\varphi_{d,k}}	\label{eq:b2_diff_mapping}
\end{align}
The common part of the two spectral coefficients is given by $\Delta t_k$ for magnitude and $\varphi_{c,k}$ for phase. The relative magnitude information is determined by $\theta_k$ whereas $\varphi_{d,k}$ represents the differential phase. In this representation, $\Delta t_k$ and $\theta_k$ do conceptually correspond to the temporal shift and the polarization angle of the corresponding solitonic component.

\subsection{Estimation}\label{sec:estimation}

The differential information $\theta_k$ and $\varphi_{d,k}$ can be detected from \eqref{eq:spec_amp_ratio} without being (much) affected by the eigenvalue fluctuations. We have
\begin{align}\label{eq:diff_estimation}
\hat{\theta}_k & =\arctan \, |\hat{b}_d(\lambda_k)|	\nonumber
\\
\hat{\varphi}_{d,k} & =\arg \, \hat{b}_d(\lambda_k)
\end{align}
with
\begin{equation}\label{eq:bd_est}
\hat{b}_d(\lambda_k)=\frac{b_2\left(\lambda_k(L),L\right)}{b_1\left(\lambda_k(L),L\right)}.
\end{equation}

The common part, however, undergoes the perturbed transformation affected by $\lambda_k(z)$ and needs to be equalized. We use the MBR \eqref{eq:practical_inv}, but any other equalizer, e.g. MMSE, can be used. The common part is obtained by
\begin{align}\label{eq:common_estimation}
\hat{\Delta t} & =\frac{1}{4 \sigma} \ln \left(\left|\hat{b}_{1,\mathrm{MBR}}(\lambda_k)\right|^2+\left|\hat{b}_{2,\mathrm{MBR}}(\lambda_k)\right|^2\right)	\nonumber
\\
\hat{\varphi}_{c,k} & =\arg \, \left\{ \hat{b}_{1,\mathrm{MBR}}(\lambda_k) \right\}
\end{align}
with $\hat{b}_{i,\mathrm{MBR}}(\lambda_k)$ according to \eqref{eq:practical_inv}.

\section{Simulation Results}\label{sec:simulation}

We evaluated our precoding scheme in SSFM simulations for different modulation schemes and compared it to the conventional independent modulation of the spectral coefficients. The block diagram of the system is shown in Fig.~\ref{fig:trans_setup}.

For the conventional standard approach, the magnitudes and phases of the spectral coefficients $b_1(\lambda_k)$ and $b_2(\lambda_k)$ are directly and independently modulated. The time domain signal is generated via Alg.~\ref{alg:DT2}. After calculating the nonlinear spectrum via NFT at the receiver, the spectral coefficients are equalized independently by the MBR according to \eqref{eq:practical_inv}. In the following, to simplify notation, we always assume $\hat{b}_i(\lambda_k) \coloneqq \hat{b}_{i,\mathrm{MBR}}$ and denote magnitude and phase as $|\hat{b}_{i}(\lambda_k)|$ and ${\hat{\varphi}_{i,k}=\arg \, \hat{b}_{i}(\lambda_k)}$.

When applying the precoding, we modulate $\Delta t_k$, $\theta_k$, $\varphi_{c,k}$ and $\varphi_{d,k}$. The corresponding spectral coefficients are used to generate the time domain signal via Alg.~\ref{alg:DT2}. At the receiver, the NFT is applied to the received signal for transformation into the nonlinear spectral domain. The estimation of the differential information $\hat{\varphi}_{d,k}$ and $\hat{\theta}_k$ is done according to \eqref{eq:diff_estimation} and \eqref{eq:bd_est}. For estimating the common information $\hat{\varphi}_{c,k}$ and $\hat{\Delta t}$ according to \eqref{eq:common_estimation}, we still use the MBR equalization.

We consider two transmission scenarios: First, we present simulation results for first order solitons where we apply the precoding to the magnitudes and phases of the spectral coefficients. Our precoding scheme is generally applicable to multi-solitons. Therefore we secondly adapt our simulation to the experimental second order soliton transmission scheme in \cite{Gaiarin2018}, where only phases of $b_1(\lambda_k)$ and $b_2(\lambda_k)$ are independently QPSK modulated.

In Sec.~\ref{sec:motivation} and \ref{sec:precoding}, our precoding was motivated and described based on the simplified concept of (ideal) distributed Raman amplification. However, lumped amplification using EDFAs is commonly used in practice. In that case, due to the fiber attenuation, the eigenvalues do not remain constant in between two amplifiers where the noise is injected. Nevertheless, a piecewise model similar to Fig.~\ref{fig:distRaman_model} is still applicable, where fiber attenuation adds perturbation to the eigenvalues and spectral coefficients. A usual method is the lossless path-averaged model which assumes a fixed path-averaged eigenvalue in each span. We consider such a link and simulate the (experimental) setup in \cite{Gaiarin2018}. 

The fiber parameters are chosen as: nonlinearity coefficient $\gamma=1.25\, \frac{1}{\mathrm{W\, km}}$, dispersion coefficient $\beta_2=-21.67\, \frac{\mathrm{ps}^2}{\mathrm{km}}$ and attenuation coefficient $\alpha=0.0459\, \frac{1}{\mathrm{km}}$. For amplification, EDFAs are considered with $\ell_\mathrm{span}=41.5 \, \mathrm{km}$ amplifier spacing. Being aware of the generally larger distortion in a practical experiment compared to our simplified simulation model, we increase the total propagation distance in our simulation by factor $8$ to $2988\mathrm{km}$ and assume an EDFA noise figure of $F=10\mathrm{dB}$. The time normalization is $T_0=47\mathrm{ps}$. The effective nonlinearity coefficient $\gamma_{\mathrm{eff}}=\gamma (1-\exp(-\alpha \ell_\mathrm{span}))/(\alpha \ell_\mathrm{span})$ of the lossless path-averaged model is used for power normalization according to \eqref{eq:Q_physical}. We use a simulation time frame that is $10$ times larger than the pulse duration $T$ (defined to contain $99.99\%$ of the total soliton energy) and apply a windowing function in the SSFM simulation after each EDFA noise injection, to avoid simulation errors arising from high bandwidth noise components leaving the simulation time window. To keep numerical errors in the NFT calculation small, we truncate the signal at $1.5\, T$. Shorter truncation windows induce further perturbations.

We calculate the nonlinear discrete spectrum $\{\lambda_k(L),b_1(\lambda_k(L),L),b_2(\lambda_k(L),L)\}$ at $9$ different locations along the link (every $332\, \mathrm{km}$) using the dual polarization extension of the Forward-Backward method in~\cite{aref2016control}. A lowpass filter with a cutoff frequency matched to the maximum occurring soliton pulse bandwidth in the respective scenario is applied before signal normalization and calculation of the nonlinear spectrum via the NFT.

\subsection{First Order Soliton Modulation}

We consider the spectral coefficient phase and magnitude modulation for first order solitons with $\lambda_1=0.5j$. Since there is only a single eigenvalue $\lambda_1$, we drop the index $k$ in this subsection and define ${b_i \coloneqq b_i(\lambda_1)=|b_i|\exp(j\varphi_i)}$. We consider the two scenarios of conventional independent modulation of $b_1$ and $b_2$ and our precoding approach separately in Subsec.~\ref{sec:1Sol_indepMod} and Subsec.~\ref{sec:1Sol_diffMod} and compare them in Subsec.~\ref{sec:1Sol_comparison}. Note that the representation \eqref{eq:b1_com_mapping}~-~\eqref{eq:b2_diff_mapping} allows a simple relation to the time domain signal. First order soliton pulses with eigenvalue $\lambda=\omega + j\sigma$ can be expressed as
\begin{align}\label{eq:1Soliton_2Pol}
q_1(t,z)=&2\sigma |\cos(\theta)| \mathrm{sech}\left(2\sigma\left(t-\Delta t + 4\omega z\right)\right)
\cdot \exp(-j\varphi_1-2j\omega t -4j(\omega^2-\sigma^2)z)\	\nonumber
\\
q_2(t,z)=&2\sigma |\sin(\theta)| \mathrm{sech}\left(2\sigma\left(t-\Delta t + 4\omega z\right)\right)
\cdot \exp(-j\varphi_2-2j\omega t -4j(\omega^2-\sigma^2)z)
\end{align} 

\subsubsection{Independent Modulation of Spectral Coefficients}\label{sec:1Sol_indepMod}

For the conventional reference system, the phases $\varphi_i$ of $b_i$ are modulated independently by an identical 8-PSK constellation. The magnitudes $|b_1|$ and $|b_2|$ are independently modulated as well using the following constellation: 
\begin{equation*}
|b_i|\in \left\{\frac{\exp(-1)}{\sqrt{2}},\frac{1}{\sqrt{2}},\frac{\exp(1)}{\sqrt{2}}\right\}.
\end{equation*} 
The reasons for choosing this symbol alphabet are as follows:
\begin{itemize}
\item Modulating the magnitudes of the spectral coefficients corresponds to a temporal pulse shift in time domain. According to \eqref{eq:b1_diff_mapping}, \eqref{eq:b2_diff_mapping} and \eqref{eq:1Soliton_2Pol}, this constellation choice ensures the maximum temporal shift of the pulse to be $\Delta t=\pm 1$ which defines the pulse time frame.
\item As it will be seen later, it is beneficial to perform the symbol decision based on $\ln \left( |b_i| \right)$. In log-domain, the above constellation points are equidistant. 
\end{itemize}

\begin{figure}
\center
\includegraphics[scale=0.3]{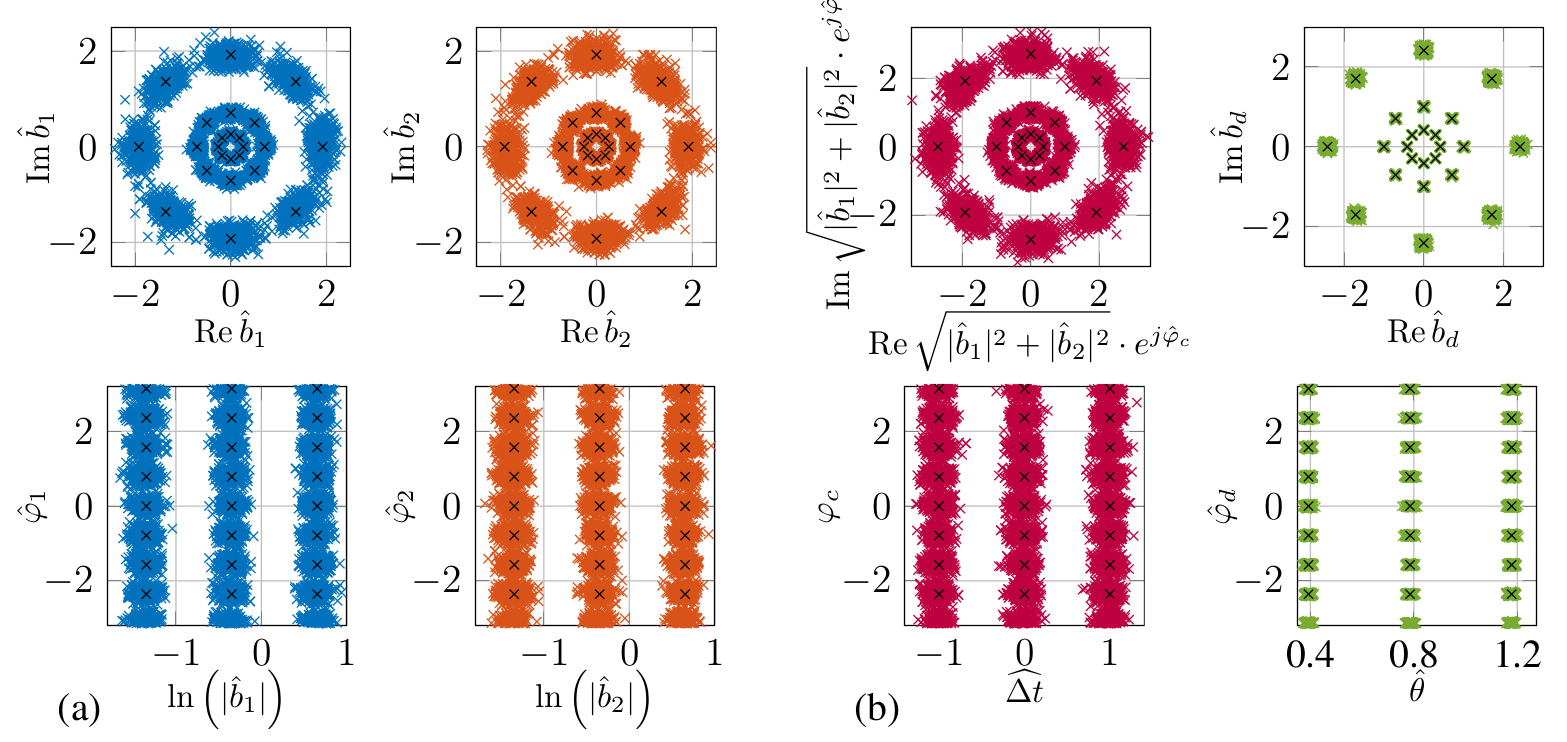}
\caption{Estimated constellations for first order soliton spectral coefficients at $L\cdot L_0=2988\mathrm{km}$: (a) Independent modulation with estimation via MBR \eqref{eq:practical_inv} and (b) differential modulation with estimation via \eqref{eq:diff_estimation} and \eqref{eq:common_estimation}.}
\label{fig:RxConst}
\end{figure}

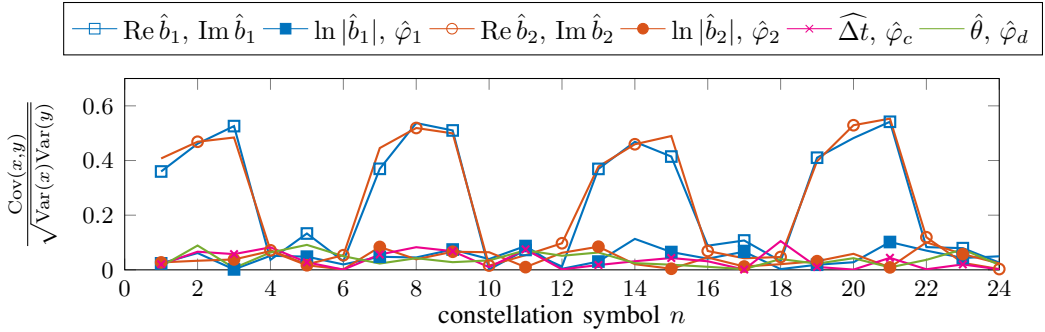
\begin{figure}
\center
\begin{tikzpicture}[baseline=(current axis.south)]
\begin{axis}[ticklabel style={font=\footnotesize},width=0.8\columnwidth,height=0.25\textwidth,xmin=0,xmax=24,ymin=0,ymax=0.7,label style={font=\small},ylabel={$\frac{\mathrm{Cov}(x,y)}{\sqrt{\mathrm{Var}(x) \mathrm{Var}(y)}}$},xlabel={constellation symbol $n$},x label style={yshift=0.5em},legend style={font=\small},title style={font=\small},legend columns=6,legend style={at={(1.05,1.4)}}]

\addplot[forget plot,thick,mark=square,myblue,mark repeat=2,width=\linewidth]table {1Sol_independent_b1_CorrCoef_re_im.dat};
\addplot[forget plot,thick,myblue,mark=square*,mark repeat=2,width=\linewidth]table {1Sol_independent_b1_CorrCoef_abs_angle.dat};
\addplot[forget plot,thick,myred,mark=o,mark repeat=2,mark phase=2,width=\linewidth]table {1Sol_independent_b2_CorrCoef_re_im.dat};
\addplot[forget plot,thick,myred,mark=*,mark repeat=2,width=\linewidth]table {1Sol_independent_b2_CorrCoef_abs_angle.dat};
\addplot[forget plot,thick,magenta,mark=x,mark repeat=2,width=\linewidth]table {1Sol_differential_CorrCoef_deltaT_phic.dat};
\addplot[forget plot,thick,mygreen,width=\linewidth]table {1Sol_differential_CorrCoef_theta_phid.dat};

\addlegendimage{line legend,myblue,mark=square}
\addlegendentry{$\mathrm{Re} \, \hat{b}_1, \,  \mathrm{Im}\, \hat{b}_1$}
\addlegendimage{line legend,myblue,mark=square*}
\addlegendentry{$\ln|\hat{b}_1|, \, \hat{\varphi}_1$}
\addlegendimage{line legend,myred,mark=o}
\addlegendentry{$\mathrm{Re} \, \hat{b}_2, \, \mathrm{Im}\, \hat{b}_2$}
\addlegendimage{line legend,myred,mark=*}
\addlegendentry{$\ln|\hat{b}_2|, \, \hat{\varphi}_2$}
\addlegendimage{line legend,magenta,mark=x}
\addlegendentry{$\widehat{\Delta t}, \, \hat{\varphi}_c$}
\addlegendimage{line legend,mygreen}
\addlegendentry{$\hat{\theta}, \, \hat{\varphi}_d$}

\end{axis}
\end{tikzpicture}
\caption{Independent modulation: Correlation coefficient of real and imaginary part or phase and logarithmic magnitude of $\hat{b}_1$ and $\hat{b}_2$. Differential Modulation: Correlation coefficient of differential ($\hat{\theta}$, $\hat{\varphi}_d$) and common ($\hat{\Delta t}$, $\hat{\varphi}_c$) information}
\label{fig:corr_coef_1Sol}
\end{figure}

\begin{figure}
\center
\begin{tikzpicture}[baseline=(current axis.south)]
\begin{axis}[axis y line*=left,ticklabel style={font=\footnotesize},width=0.8\columnwidth,height=0.25\textwidth,xmin=-1.8,xmax=1.3,ymin=0,ymax=1.8,y label style={yshift=-1em},label style={font=\small},ylabel={Empirical PDF},xlabel={$\ln \left( |\hat{b}_i| \right)$},x label style={yshift=0.5em},legend style={font=\small},title style={font=\small},legend columns=2,legend style={at={(1.02,1.8)}}]

\addplot[ybar interval,forget plot,thick,myblue,width=\linewidth]table {1Sol_indep_hist_b1.dat};
\addplot[forget plot,thick,black,width=\linewidth]table {1Sol_indep_gaussApprox1_b1.dat};
\addplot[forget plot,thick,black,width=\linewidth]table {1Sol_indep_gaussApprox2_b1.dat};
\addplot[forget plot,thick,black,width=\linewidth]table {1Sol_indep_gaussApprox3_b1.dat};

\end{axis}
\begin{axis}[axis y line*=right,xticklabels={,,},ticklabel style={font=\footnotesize},width=0.8\columnwidth,height=0.25\textwidth,xmin=-1.8,xmax=1.3,ymin=0,ymax=1,y label style={yshift=-1em},label style={font=\small},x label style={yshift=0.5em},legend style={font=\small},title style={font=\small},legend columns=2,legend style={at={(1.02,1.8)}}]

\addplot[forget plot,thick,myred,width=\linewidth,each nth point=50]table {1Sol_indep_CDF_b1.dat};

\end{axis}
\end{tikzpicture}

\caption{Histograms and respectively fitted Gaussian distributions for estimated constellations of $\ln |\hat{b}_1|$ and $\ln |\hat{b}_2|$ for the classical independent modulation scenario at $L \cdot L_0=2988 \mathrm{km}$}
\label{fig:noise_histograms_indep}
\end{figure}
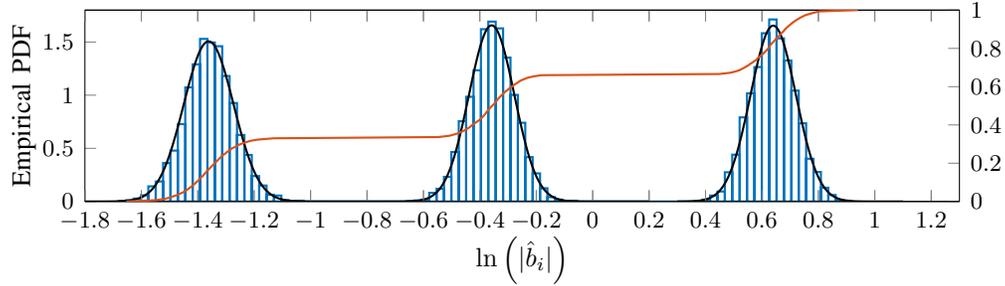

We generate and transmit $40000$ soliton signals where spectral coefficients are modulated using the constellations for phase and magnitude given above. Fig.~\ref{fig:RxConst}~(a) shows the estimated constellations of $\hat{b}_i$ via the MBR from the received $b_i(\lambda(L),L)$ at the end of the link at $L\cdot L_0=2988\mathrm{km}$. We also show $\ln(\hat{b}_i)=\ln(|\hat{b}_i| \exp(j \hat{\varphi}_i))=\ln \left( |\hat{b}_i| \right) +j \hat{\varphi}_i$ which we prefer to consider for evaluation and symbol decision due to the following reasons:
First, phase and magnitude of $\hat{b}_i$ are much less correlated than real and imaginary part of $\hat{b}_i$. The respective (linear) correlation coefficients are shown in Fig.~\ref{fig:corr_coef_1Sol}, separately for all $24$ constellation points.
Second, considering the histogram of the estimated $\ln \left( |\hat{b}_i| \right)$ in addition with respectively fitted Gaussian distributions in Fig.~\ref{fig:noise_histograms_indep}, one can observe that the noise distribution is (almost) independent of the constellation point (both $\ln \left( |\hat{b}_1| \right)$ and $\ln \left( |\hat{b}_2| \right)$ behave identical). Remember, that we have chosen the constellation points to be equidistant in terms of $\ln \left( |b_i| \right)$.

\subsubsection{Differential Modulation of Spectral Coefficients}\label{sec:1Sol_diffMod}

Now we consider the differential modulation of the spectral coefficients with encoding according to Sec.~\ref{sec:Encoding}. Thus, we independently modulate $\{\Delta t,\, \theta, \,\varphi_c, \, \varphi_d\}$. The magnitude of the spectral coefficients $|b_1|,\, |b_2|$ is determined by $\Delta t,\, \theta$ while the phase of the spectral coefficients is determined by $\varphi_c$ and $\varphi_d$ determine the phase information. 
For a fair comparison of independent and differential modulation, both scenarios should have the same number of constellation points for each degree of freedom. In addition, the mean signal power for the transmission pulses should be the same.
To achieve that, the phases $\varphi_c$ and $\varphi_d$ are chosen from the same $8$-PSK constellations as before while $\Delta t$ and $\theta$ are chosen from the following constellations

\begin{align*}
\Delta t \in \left\{-1,0,1\right\},
\qquad \qquad
\theta \in \left\{\frac{\pi}{8},\frac{\pi}{4},\frac{3\pi}{8}\right\}.
\end{align*}

These constellations guarantee (i) the symmetry between $b_1(\lambda_k)$ and $b_2(\lambda_k)$ and (ii) the identical power and identical pulse duration compared to the former modulation case. Recall that $\Delta t$ is the temporal shift of the soliton pulse according to \eqref{eq:1Soliton_2Pol}. We choose equidistant constellation points for both $\Delta t$ and $\theta$ as the variances of the received $\hat{\Delta t}$ and $\hat{\theta}$ appear to be mean independent (see also~\cite{Gordon1986}). We discuss this observation later on.

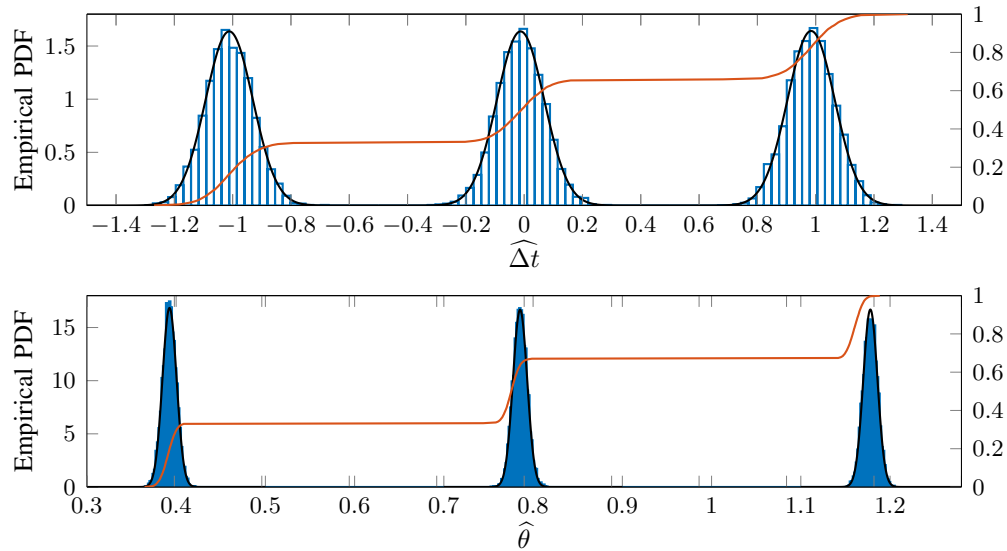
\begin{figure}
\center
\begin{tikzpicture}[baseline=(current axis.south)]
\begin{axis}[axis y line*=left,ticklabel style={font=\footnotesize},width=0.8\columnwidth,height=0.25\textwidth,xmin=-1.5,xmax=1.5,ymin=0,ymax=1.8,y label style={yshift=-1em},label style={font=\small},ylabel={Empirical PDF},xlabel={$\widehat{\Delta t}$},x label style={yshift=0.5em},legend style={font=\small},title style={font=\small},legend columns=2,legend style={at={(1.02,1.8)}}]

\addplot[ybar interval,forget plot,thick,myblue,width=\linewidth]table {1Sol_diff_hist_deltaT.dat};
\addplot[forget plot,thick,black,width=\linewidth]table {1Sol_diff_gaussApprox1_deltaT.dat};
\addplot[forget plot,thick,black,width=\linewidth]table {1Sol_diff_gaussApprox2_deltaT.dat};
\addplot[forget plot,thick,black,width=\linewidth]table {1Sol_diff_gaussApprox3_deltaT.dat};

\end{axis}
\begin{axis}[axis y line*=right,xticklabels={,,},ticklabel style={font=\footnotesize},width=0.8\columnwidth,height=0.25\textwidth,xmin=-1.5,xmax=1.5,ymin=0,ymax=1,y label style={yshift=-1em},label style={font=\small},x label style={yshift=0.5em},legend style={font=\small},title style={font=\small},legend columns=2,legend style={at={(1.02,1.8)}}]

\addplot[forget plot,thick,myred,width=\linewidth,each nth point=50]table {1Sol_diff_CDF_deltaT.dat};

\end{axis}
\end{tikzpicture}
\begin{tikzpicture}[baseline=(current axis.south)]
\begin{axis}[axis y line*=left,ticklabel style={font=\footnotesize},width=0.8\columnwidth,height=0.25\textwidth,xmin=0.3,xmax=1.28,ymin=0,ymax=18,y label style={yshift=-1em},label style={font=\small,align=center},ylabel={Empirical PDF},xlabel={$\widehat{\theta}$},x label style={yshift=0.5em},legend style={font=\small},title style={font=\small},legend columns=2,legend style={at={(1.02,1.8)}}]

\addplot[ybar interval,forget plot,thick,myblue,width=\linewidth]table {1Sol_diff_hist_theta.dat};
\addplot[forget plot,thick,black,width=\linewidth]table {1Sol_diff_gaussApprox1_theta.dat};
\addplot[forget plot,thick,black,width=\linewidth]table {1Sol_diff_gaussApprox2_theta.dat};
\addplot[forget plot,thick,black,width=\linewidth]table {1Sol_diff_gaussApprox3_theta.dat};

\end{axis}
\begin{axis}[axis y line*=right,xticklabels={,,},ticklabel style={font=\footnotesize},width=0.8\columnwidth,height=0.25\textwidth,xmin=0.3,xmax=1.3,ymin=0,ymax=1,y label style={yshift=-1em},label style={font=\small},x label style={yshift=0.5em},legend style={font=\small},title style={font=\small},legend columns=2,legend style={at={(1.02,1.8)}}]

\addplot[forget plot,thick,myred,width=\linewidth,each nth point=50]table {1Sol_diff_CDF_theta.dat};

\end{axis}
\end{tikzpicture}
\caption{Histograms and respectively fitted Gaussian distributions for estimated constellations of $\widehat{\Delta t}$ and $\hat{\theta}$ at $L\cdot L_0 =2988\mathrm	{km}$ for the differential modulation scenario}
\label{fig:noise_histograms_diff}
\end{figure}

We generate $40000$ soliton signals where the spectral coefficients are calculated from the randomly chosen $\{\Delta t, \, \theta, \, \varphi_c, \, \varphi_d\}$. Fig.~\ref{fig:RxConst}~(b) illustrates the estimated constellations after propagation of $L\cdot L_0=2988 \mathrm{km}$. Like in Fig.~\ref{fig:RxConst}~(a), we present the same received constellations by two representations: in polar coordinates in terms of magnitude and phase and in log-domain in terms of ${\hat{\Delta t}, \, \hat{\varphi}_c, \, \hat{\theta}, \, \hat{\varphi}_d}$. The latter representation is beneficial for the two following reasons:\\
(i) We illustrate in Fig.~\ref{fig:noise_histograms_diff} that the distributions of the estimated $\hat{\Delta t}$ around the respective constellation points is almost identical. The same is true for $\hat{\theta}$.\\
(ii) There is very small correlation between  $\widehat{\Delta t}$ and $\hat{\varphi}_c$ as well as $\hat{\theta}$ and $\hat{\varphi}_d$. This is shown in Fig.~\ref{fig:corr_coef_1Sol}.
None of these properties holds between magnitude and phase in the polar coordinate representation.

\subsubsection{Comparison}\label{sec:1Sol_comparison}

\begin{figure}
\center
\begin{tikzpicture}[baseline=(current axis.south)]
\begin{axis}[ymode=log,ticklabel style={font=\footnotesize},width=0.8\columnwidth,height=0.25\textwidth,xmin=1.6284,xmax=15,ymin=1e-5,ymax=0.05,label style={font=\small},ylabel={$Var(\hat{\varphi}-\varphi)$},xlabel={$L$},x label style={yshift=0.5em},legend style={font=\small,legend pos=south east},legend columns=4,title style={font=\small}]

\addplot[forget plot,thick,mark=square,mark repeat=2,myblue,width=\linewidth]table {1Sol_independent_arg_b1_variance.dat};
\addplot[forget plot,thick,myred,mark=*,mark phase=2,mark repeat=2,width=\linewidth]table {1Sol_independent_arg_b2_variance.dat};
\addplot[forget plot,thick,purple,mark=x,width=\linewidth]table {1Sol_differential_phi_c_variance.dat};
\addplot[forget plot,thick,mygreen,mark=diamond,width=\linewidth]table {1Sol_differential_phi_d_variance.dat};

\addlegendimage{line legend,myblue,mark=square}
\addlegendentry{$\varphi_1$}
\addlegendimage{line legend,myred,mark=*}
\addlegendentry{$\varphi_2$}
\addlegendimage{line legend,purple,mark=x}
\addlegendentry{$\varphi_c$}
\addlegendimage{line legend,mygreen,mark=diamond}
\addlegendentry{$\varphi_d$}

\end{axis}
\node at (-0.3,-0.7) {(a)};
\end{tikzpicture}
\hspace{5mm}
\begin{tikzpicture}[baseline=(current axis.south)]
\begin{axis}[ymode=log,ticklabel style={font=\footnotesize},width=0.8\columnwidth,height=0.25\textwidth,xmin=1.6284,xmax=15,ymin=1e-5,ymax=0.01,label style={font=\small},ylabel={$\frac{Var(\hat{x}-x)}{d^2}$},xlabel={$L$},x label style={yshift=0.5em},legend style={font=\small,legend pos=south east},legend columns=4,title style={font=\small}]

\addplot[forget plot,thick,mark=square,mark repeat=2,myblue,width=\linewidth]table {1Sol_independent_ln_abs_b1_variance.dat};
\addplot[forget plot,thick,myred,mark=*,mark phase=2,mark repeat=2,width=\linewidth]table {1Sol_independent_ln_abs_b2_variance.dat};
\addplot[forget plot,thick,purple,mark=x,width=\linewidth]table {1Sol_differential_deltaT_variance.dat};
\addplot[forget plot,thick,mygreen,mark=diamond,width=\linewidth]table [x expr=\thisrowno{0},y expr=\thisrowno{1}/0.1542] {1Sol_differential_theta_variance.dat};

\addlegendimage{line legend,myblue,mark=square}
\addlegendentry{$\ln \left( |b_1| \right)$}
\addlegendimage{line legend,myred,mark=*}
\addlegendentry{$\ln \left( |b_1| \right)$}
\addlegendimage{line legend,purple,mark=x}
\addlegendentry{$\Delta t$}
\addlegendimage{line legend,mygreen,mark=diamond}
\addlegendentry{$\theta$}

\end{axis}
\node at (-0.3,-0.7) {(b)};
\end{tikzpicture}
\caption{Comparison of independent and differential modulation: Variances for estimation error of phase (a) and magnitude (b) information along different positions along the link $L$. The normalized unit $L=1$ corresponds to approximately $204\mathrm{km}$.}
\label{fig:estiamtionErrorVariance}
\end{figure}
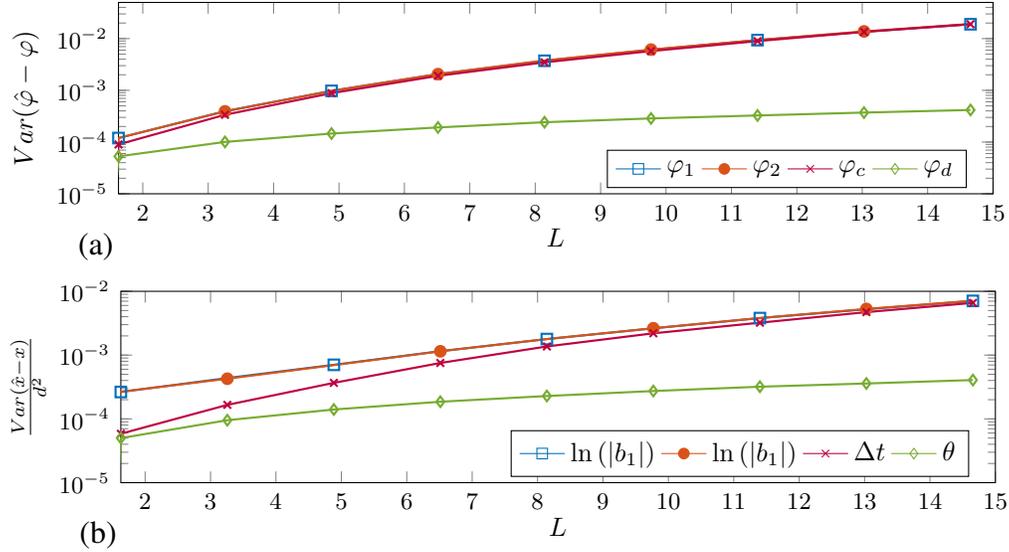

The advantage of differential modulation over independent modulation is rather clear from Fig.~\ref{fig:RxConst}. The differential part is much less distorted than the common part which is received almost as erroneous as the independently modulated $b_1$ and $b_2$. To quantify these observations, we compare the variances of their estimation errors in Fig.~\ref{fig:estiamtionErrorVariance}. They are calculated at $9$ locations along the link from the transmission of 40000 randomly modulated solitons. Note that all variances are normalized by the squared constellation point spacing $d^2$ ($d_{\theta}=\frac{\pi}{8}$, $d_{\Delta t}=d_{\ln |b_i|}=1$). 
\footnote{When the received constellation points, like in our problem, have a Gaussian-like distribution, the symbol error rate can be approximated simply by applying the normal Q-function to the inverse of the square root normalized variance.}
We observe that the differential part, $\hat{\theta}$ and $\hat{\varphi_d}$, has a much smaller variance and is thus much less erroneous than the common part $\hat{\Delta t}$ and $\hat{\varphi_c}$. The variance of $\hat{\varphi_c}$ is almost the same as the variances of $\hat{\varphi}_1$ and $\hat{\varphi}_2$ in the independent modulation case. However, it is interesting to see that the variance of $\widehat{\Delta t}$ is smaller than the variance of $\ln |\hat{b}_i|$ for small $L$ but they get closer for increasing $L$. With the aid of simulations, our hand-waving explanation is that we can generally write
\begin{equation*}
\ln \left( |\hat{b}_1| \right)=2\sigma \widehat{\Delta t} + \ln \left( |\cos \hat{\theta}| \right).
\end{equation*}
We have observed that $\widehat{\Delta t}$ and $\hat{\theta}$ have small correlation along the link, see e.g. Fig.~\ref{fig:corr_coef_1Sol}. If we assume that they are uncorrelated (which may not be the case rigorously), then the variance of $\ln \left( |\hat{b}_1| \right)$ is the sum of the variances of $2\sigma \widehat{\Delta t}$ and $\ln |\cos \hat{\theta} |$. The variance of $\widehat{\Delta t}$ and $\ln \left( |\hat{b}_1| \right)$ get similar when the variance of $\widehat{\Delta t}$ is much larger than the one of $\hat{\theta}$. This happens for long link lengths (compare Fig.~\ref{fig:estiamtionErrorVariance}~(b)).

\subsection{Phase Modulation for Second Order Soliton}

Now we consider second order soliton phase modulation by mimicking the experiment in \cite{Gaiarin2018}. We use the same second order soliton ($\lambda_1=0.3j$ and $\lambda_2=0.6j$) QPSK modulation of $b_i(\lambda_k)$. The QPSK constellation of $b_i(\lambda_2)$ is chosen such that it is rotated by $\pi/4$ compared to the QPSK constellation of $b_i(\lambda_1)$. Both phases are modulated independently. The magnitudes are fixed to $|b_i(\lambda_1)|=0.14$ and $|b_i(\lambda_2)|=5$. The resulting soliton's pulse duration is approximately $T \approx 1.05 \mathrm{ns}$.

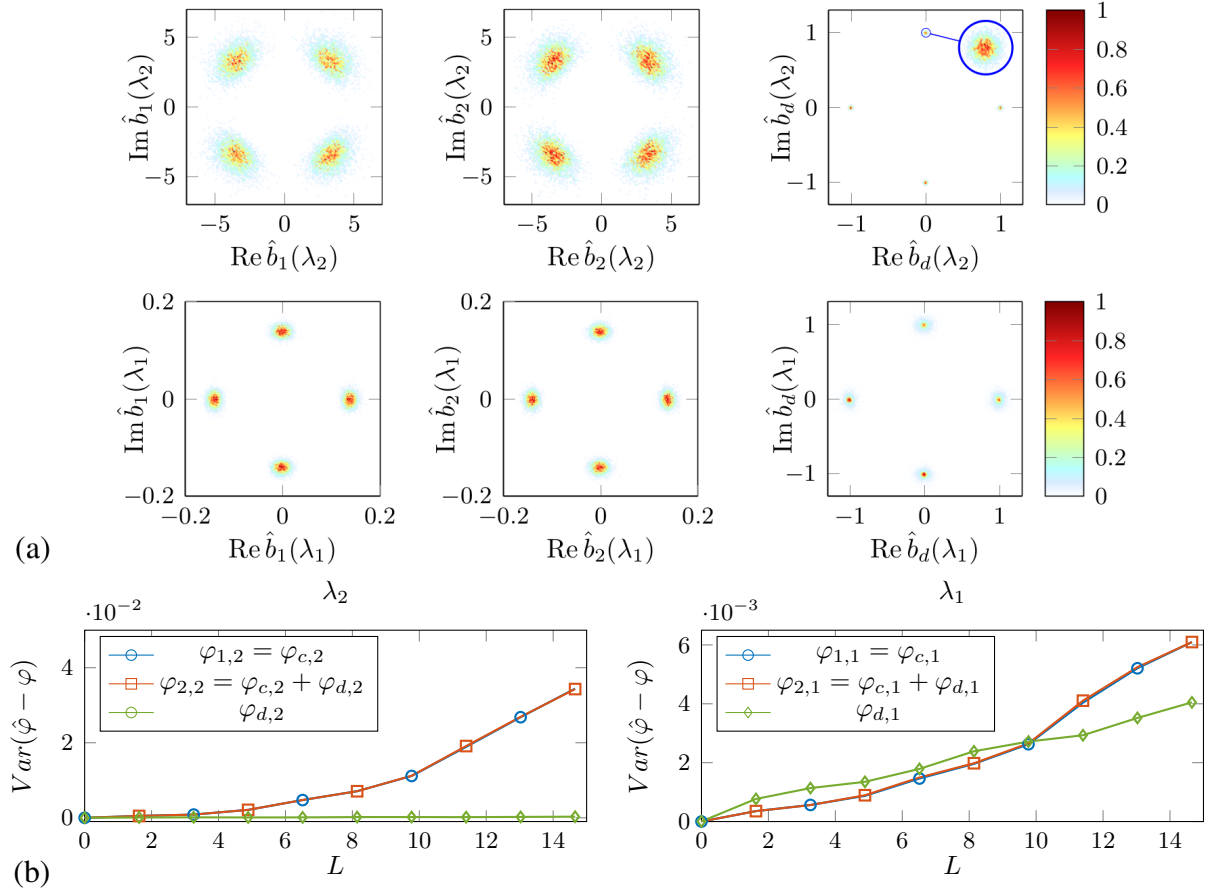
\begin{figure*}
\includegraphics[scale=0.255]{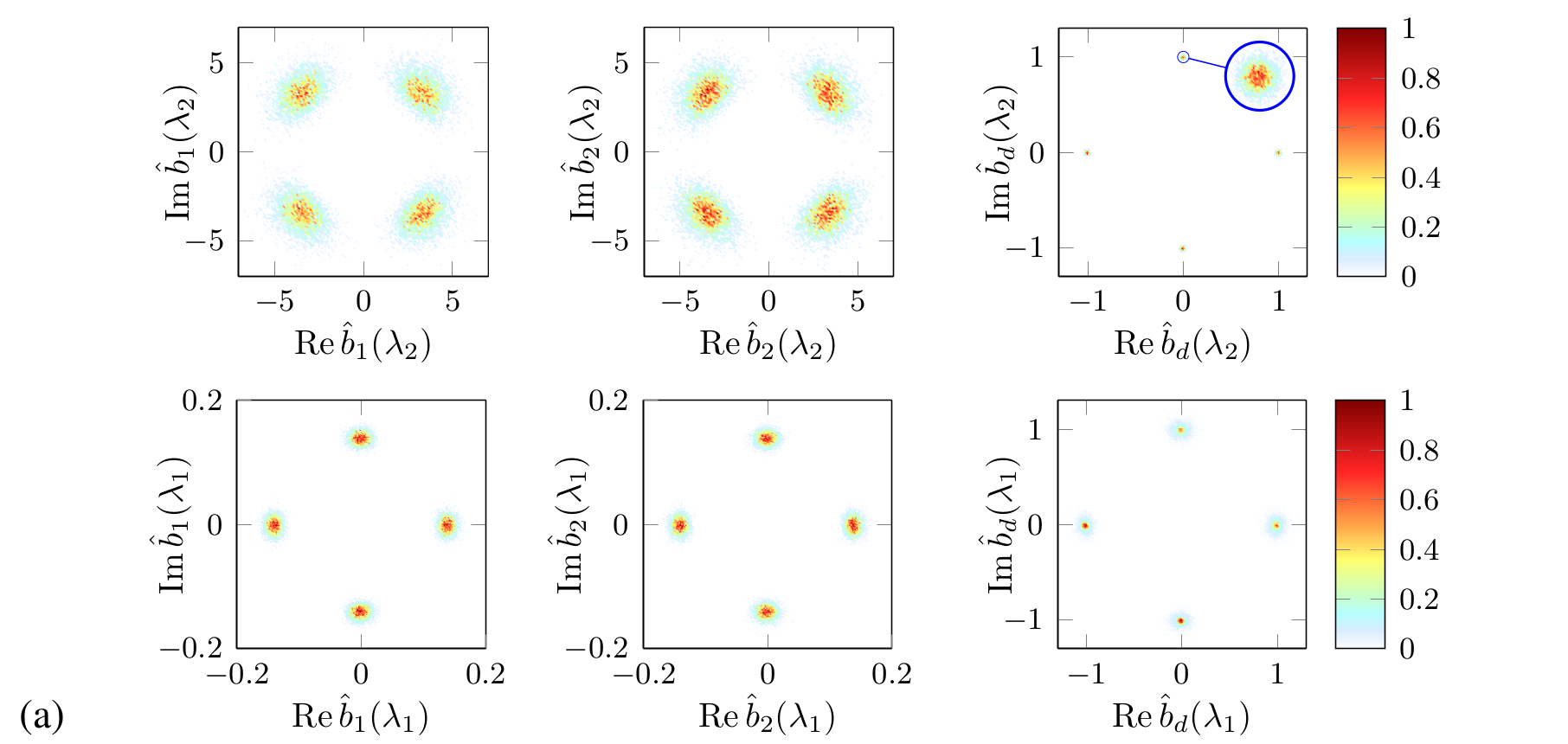}

\begin{tikzpicture}[baseline=(current axis.south)]
\begin{axis}[ticklabel style={font=\footnotesize},width=0.5\textwidth,height=0.25\textwidth,xmin=0,xmax=15,ymin=-0.001,ymax=0.05,y label style={yshift=-1em},label style={font=\small},ylabel={$Var(\hat{\varphi}-\varphi)$},xlabel={$L$},x label style={yshift=0.5em},legend style={font=\small,legend pos=north west},title style={font=\small},title={$\lambda_2$}]

\addplot[forget plot,thick,myblue,mark repeat=2,mark=o,width=\linewidth]table {2Sol_ldv2_b1_phaseVariance.dat};
\addplot[forget plot,thick,myred,mark repeat=2,mark phase=2,mark=square,width=\linewidth]table {2Sol_ldv2_b2_phaseVariance.dat};
\addplot[forget plot,thick,mygreen,mark=diamond,width=\linewidth]table {2Sol_ldv2_bd_phaseVariance.dat};

\addlegendimage{line legend,myblue,mark=o}
\addlegendentry{$\varphi_{1,2}=\varphi_{c,2}$}
\addlegendimage{line legend,myred,mark=square}
\addlegendentry{$\varphi_{2,2}=\varphi_{c,2}+\varphi_{d,2}$}
\addlegendimage{line legend,mygreen,mark=o}
\addlegendentry{$\varphi_{d,2}$}

\end{axis}
\node at (-0.7,-0.7) {(b)};
\end{tikzpicture}
\hspace{1mm}
\begin{tikzpicture}[baseline=(current axis.south)]
\begin{axis}[ticklabel style={font=\footnotesize},width=0.5\textwidth,height=0.25\textwidth,xmin=0,xmax=15,ymin=0,ymax=0.0065,y label style={yshift=-1em},label style={font=\small},ylabel={$Var(\hat{\varphi}-\varphi)$},xlabel={$L$},x label style={yshift=0.5em},legend style={font=\small,legend pos=north west},title style={font=\small},title={$\lambda_1$}]

\addplot[forget plot,thick,myblue,mark repeat=2,mark=o,width=\linewidth]table {2Sol_ldv1_b1_phaseVariance.dat};
\addplot[forget plot,thick,myred,mark repeat=2,mark phase=2,mark=square,width=\linewidth]table {2Sol_ldv1_b2_phaseVariance.dat};
\addplot[forget plot,thick,mygreen,mark=diamond,width=\linewidth]table {2Sol_ldv1_bd_phaseVariance.dat};

\addlegendimage{line legend,myblue,mark=o}
\addlegendentry{$\varphi_{1,1}=\varphi_{c,1}$}
\addlegendimage{line legend,myred,mark=square}
\addlegendentry{$\varphi_{2,1}=\varphi_{c,1}+\varphi_{d,1}$}
\addlegendimage{line legend,mygreen,mark=diamond}
\addlegendentry{$\varphi_{d,1}$}

\end{axis}

\end{tikzpicture}

\caption{(a) Empirical PDF for the estimated spectral coefficients $\hat{b}_i(\lambda_1)$ and $\hat{b}_i(\lambda_2)$ and the differential information $\hat{b}_d(\lambda_k)$. (b) Variance of phase estimation error for detection of $\hat{\varphi}_{1,k}$ and $\hat{\varphi}_{2,k}$ via MBR \eqref{eq:practical_inv} and $\hat{\varphi}_{c,k}$ and $\hat{\varphi}_{d,k}$ via differential decoding \eqref{eq:diff_estimation}, \eqref{eq:common_estimation}}

\label{fig:twoSoliton_Rx_b_L}
\end{figure*}

We transmit again $40000$ randomly generated soliton pulses, where the phases $\varphi_{i,k}$ ($i,k=1,2$) of the spectral coefficients are \emph{independently} chosen from the respective QPSK constellation described above. In a PSK scenario (phase modulation only), we can directly compare the precoding with the differential detection in the same simulation with the bijective mapping $\varphi_{d,k}=\varphi_{2,k}-\varphi_{1,k}$ and $\varphi_{c,k}=\varphi_{1,k}$ according to \eqref{eq:b1_diff_mapping} and \eqref{eq:b2_diff_mapping}. 

The estimated constellations of $\hat{b}_1(\lambda_k)$, $\hat{b}_2(\lambda_k)$ and $\hat{b}_d(\lambda_k)$ according to \eqref{eq:practical_inv} and \eqref{eq:bd_est}, respectively, are shown in Fig.~\ref{fig:twoSoliton_Rx_b_L}~(a) at the end of the link $L \cdot L_0 = 2988 \mathrm{km}$. The different phase estimation error variances are compared in Fig.~\ref{fig:twoSoliton_Rx_b_L}~(b). The $\hat{b}_d(\lambda_k)$, and thus $\hat{\varphi}_{d,k}$, are much less scattered than $\hat{b}_1(\lambda_k)$ and $\hat{b}_2(\lambda_k)$ (accordingly $\hat{\varphi}_{1,k}$ and $\hat{\varphi}_{2,k}$), especially for $\lambda_2$. We also observe that the variance of $\hat{\varphi}_{d,2}$ is much smaller everywhere along the link when compared to $\hat{\varphi}_{1,2}$ and $\hat{\varphi}_{2,2}$. This is not the case for $\hat{\varphi}_{d,1}$. All $\hat{\varphi}_{1,1}$, $\hat{\varphi}_{2,1}$ and $\hat{\varphi}_{d,1}$ have rather small variance but there is no significant gain of differential precoding. To see the reason, we plot the correlation between different pairs of phases in Fig.~\ref{fig:corr_coeff_2Sol}. On one hand, we observe that the correlation between $\hat{\varphi}_{1,2}$ and $\hat{\varphi}_{2,2}$ is large. It implies that the transformation fluctuation is dominated by the eigenvalue fluctuations which leads to a small correlation between $\hat{\varphi}_{d,2}$ and $\hat{\varphi}_{c,2}$ (see Fig.~\ref{fig:corr_coeff_2Sol}). On the other hand, we observe that the correlation between $\hat{\varphi}_{1,1}$ and $\hat{\varphi}_{2,1}$ is not large. It implies that the transformation fluctuation is not dominated by the eigenvalue fluctuations. In this case, the perturbation terms $\tilde{\varepsilon}$ in \eqref{eq:spec_amp_ratio_z} can not be neglected and lead to large correlation between $\hat{\varphi}_{d,1}$ and $\hat{\varphi}_{c,1}$. The gain of the differential precoding grows when the eigenvalue fluctuations become dominant for larger propagation distances $L$.

\begin{figure}\label{fig:corr_coeff_2Sol}
\center
\begin{tikzpicture}[baseline=(current axis.south)]
\begin{axis}[ticklabel style={font=\normalsize},width=0.8\columnwidth,height=0.25\columnwidth,xmin=1.6,xmax=15,
ymin=0,ymax=1,y label style={yshift=-0.25em},label style={font=\normalsize},xlabel={$L$},x label style={yshift=0.5em},ylabel={$\frac{\mathrm{Cov}(\hat{x}-x,\hat{y}-y)}{\sqrt{\mathrm{Var}(\hat{x}-x)\mathrm{Var}(\hat{y}-y)}}$},legend style={at={(0.9,1.3)}},legend columns=4]

\addplot[myblue,forget plot,thick,mark=square]table {2Sol_corr_coef_b1b2_ldv1.dat};     
\addplot[myblue,forget plot,thick,mark=square*]table {2Sol_corr_coef_b1bd_ldv1.dat};    
\addplot[myred,forget plot,thick,mark=o]table {2Sol_corr_coef_b1b2_ldv2.dat};    
\addplot[myred,forget plot,thick,mark=*]table {2Sol_corr_coef_b1bd_ldv2.dat};

\addlegendimage{myblue,mark=square}
\addlegendentry{$\varphi_{1,1}$, $\varphi_{2,1}$}      
\addlegendimage{myblue,mark=square*}
\addlegendentry{$\varphi_{c,1}$, $\varphi_{d,1}$}   
\addlegendimage{myred,mark=o}
\addlegendentry{$\varphi_{1,2}$, $\varphi_{2,2}$}      
\addlegendimage{myred,mark=*}
\addlegendentry{$\varphi_{c,2}$, $\varphi_{d,2}$}    
    
\end{axis}

\end{tikzpicture}
\caption{Comparison of correlation coefficient between the degrees of freedom used for either independent or differential modulation for second order solitons}
\label{fig:corr_coeff_2Sol}
\end{figure}
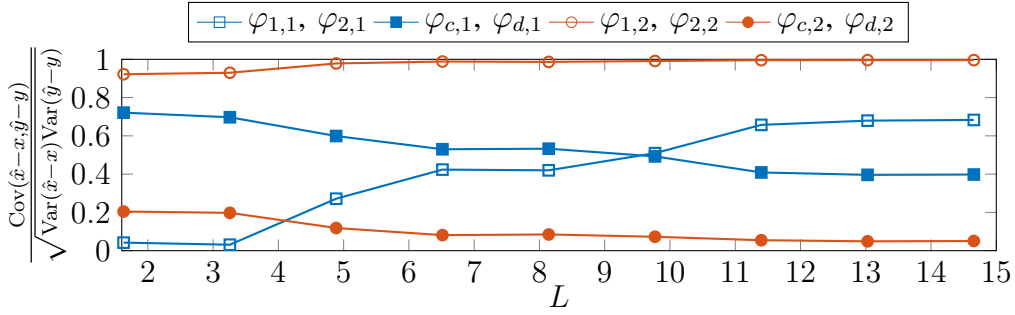

\section{Conclusion}\label{sec:conc}

We consider the transmission of dual polarization soliton pulses along a noisy fiber link. We observe that the spectral coefficients per eigenvalue $b_i(\lambda_k(z),z)$ become highly correlated along the link. We propose a precoding that maps these spectral coefficients to another set of quantities for modulation $\{ {\Delta t_k, \, \theta_k, \, \varphi_{c,k}, \, \varphi_{d,k}} \}$ with much less mutual correlation, allowing their disjoint detection with only small performance loss. In this case, the information is encoded in a common and a differential part. Since both spectral coefficients undergo a similarly perturbed (however unknown) transformation during propagation, their differential information is almost preserved. Thus, it can be recovered with high reliability directly from the received signal. Our precoding scheme thus increases the detection performance compared to the direct modulation of $b_i(\lambda_k)$ without the need of complex equalization.

We have analyzed our proposed precoding scheme and verified its benefits in SSFM simulations for first and second order solitons. The scheme is however general and can be applied to any multi-soliton pulse.

For the studied scenarios, we also showed that modulating $\ln \left(|b_i(\lambda_k)| \right)$ (as well as $\Delta t$ and $\theta$) is beneficial when compared to modulating $b_i(\lambda_k)$. That is because the received constellations of the former quantities are scattered around the transmitted symbols with much less data dependency (meaning the distribution is almost identical for all constellation points).

\ifCLASSOPTIONcaptionsoff
  \newpage
\fi

\bibliographystyle{IEEEtranTCOM}

\bibliography{references_nft}

\end{document}